\documentclass[superscriptaddress,aps,pra,twocolumn,showpacs,longbibliography]{revtex4-2}
\usepackage[utf8]{inputenc}
\usepackage{graphicx,amsmath,amsfonts,amssymb}
\usepackage{color}
\usepackage[colorlinks=true, allcolors={blue}]{hyperref}
\usepackage{graphicx}
\usepackage{epstopdf}
\usepackage{float}
\usepackage{subfigure}

\newcommand{\orcid}[1]{\href{https://orcid.org/#1}{\includegraphics[width=7pt]{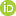}}}

\begin{document}




\title{Probing parameters estimation with Gaussian non-commutative measurements}


\author{Alice P. G. Hall}
\email{alice.hall063@academico.ufgd.edu.br}
\affiliation{Faculdade de Engenharia, Universidade Federal da Grande Dourados, Caixa Postal 364, CEP 79804-970, Dourados, MS, Brazil}

\author{Carlos H. S. Vieira}
\email{carloshsv09@gmail.com}
\affiliation{Centro de Ci\^{e}ncias Naturais e Humanas, Universidade Federal do ABC,
Avenida dos Estados 5001, 09210-580 Santo Andr\'e, S\~{a}o Paulo, Brazil}
\affiliation{Department of Physics, State Key Laboratory of Quantum Functional Materials,
and Guangdong Basic Research Center of Excellence for Quantum Science,
Southern University of Science and Technology, Shenzhen 518055, China}

\author{Jonas F. G. Santos}
\email{jonassantos@ufgd.edu.br}
\affiliation{Faculdade de Ci\^{e}ncias Exatas e Tecnologia, Universidade Federal da Grande Dourados, Caixa Postal 364, CEP 79804-970, Dourados, MS, Brazil}

\begin{abstract}
Gaussian quantum states and channels are pivotal across many branches of quantum science and their applications, including the processing and storage of quantum information, the investigation of thermodynamics in the quantum regime, and quantum computation. The great advantage is that Gaussian states are experimentally accessible via their first and second statistical moments. In this work, we investigate parameter estimation for Gaussian states, in which the probe-state preparation stage involves two noncommutative Gaussian measurements on the position and momentum observables, introducing tunable parameters. The influence of these noncommutative Gaussian measurements is investigated through the quantum Fisher information (QFI). We showed that the QFI for characterizing Gaussian channels can be increased by adjusting the uncertainty parameters in the preparation of the probe state. Furthermore, if the probe is initially in a thermal state, probe-state preparation may generate quantum coherence in its energy basis. We showed that not only does the amount of coherence affect the improvement of the QFI, but also the rate of change of the coherence with respect to the parameter to be estimated. The proposed probe-state protocol is applied to two paradigmatic single-mode Gaussian channels, the attenuator and amplification channels, which are building blocks of Gaussian quantum information. Our results contribute to the use of coherence in quantum metrology and are experimentally feasible in quantum-optical devices.  
\end{abstract}

\maketitle

\section{Introduction}

Quantum metrology aims to investigate the best possible precision for estimating physical parameters encoded within quantum systems ~\cite{PhysRevA.98.012114,giovannetti2011advances,RevModPhys.90.035005,schnabel2010quantum,toth2014quantum,haase2016precision,zwierz2012ultimate,morelli2021bayesian,seveso2017quantum,PhysRevLett.133.120601,Pritam2024QST,mukhopadhyay2024current,di2023critical,petrovnin2024microwave,degen2017quantum,montenegro2024review, Chattopadhyay2025-jk}. From a theoretical and experimental point of view, it has been proven that the Standard-Quantum Limit (SQL) sets the ultimate precision that can be achieved by using uncorrelated quantum systems \cite{Giovannetti2004-el, PhysRevLett.96.010401}. For the SQL scenario, it scales as $1/\sqrt{N}$, where $N$ is the number of uncorrelated systems. On the other hand, if non-classical resources, such as coherence and entanglement, are assumed to be available in quantum parameter estimation protocols, the sensitivity can be considerably enhanced to reach the so-called Heisenberg Limit (HL) \cite{f79z-vjsb, giovannetti2011advances}, which represents a scaling behavior of $1/N$. These limits in sensitivity have been experimentally corroborated, for instance, by estimating relative phases \cite{Daryanoosh2018-nn, PhysRevA.110.012614}. Significant recent applications of quantum metrology and sensing, ranging from gravimetry and magnetometry \cite{DeAraujo2025ws, Qvarfort2018-dr, PhysRevLett.117.138501, PhysRevResearch.7.013016}, gravitational waves detection \cite{schnabel2010quantum}, and phonon-based detection in Bose-
Einstein condensates \cite{Boixo2009-yw, Ngo2021-iy}.

Regarding continuous-variable (CV) quantum systems, the electromagnetic field in a cavity and the vibrational degree of motion of trapped ions are paradigmatic examples of quantum probes~\cite{RevModPhys.84.621, Dalvit2006-cq, Delakouras2024-lr, Santos2025}. In many relevant scenarios, the environment in which the physical parameter is to be estimated can be well described by Gaussian or non-Gaussian channels \cite{PhysRevA.104.042609}. For the former, the information about the physical parameter is completely encoded in the first and second statistical moments of the probe state, whereas for the latter, higher-order statistical moments are needed for proper estimation. Furthermore, after the interaction between the probe state and the quantum channel, the relevant question is how to estimate the optimal precision of this parameter from a sequence of probe-state measurements. This is answered by the quantum Cramér-Rao bound (QCRB)~\cite{Liu_2020}, which constitutes a lower bound to the fluctuations of a given estimator. The QCRB describes how the probe state depends on small variations in the physical parameter to be estimated, and quantum uncertainties set the bound, which is mathematically given by the quantum Fisher information (QFI).

In any parameter-estimation protocol, the probe-state preparation step is crucial. Quantum resources, such as coherence and entanglement, can be generated to enhance precision, resulting in a higher QFI for a given parameter~\cite{Giovannetti2004-el, Huang_2024}. For example, coherence on the energy basis is useful for characterizing Gaussian channels \cite{Santos2025, Frerot2024-pi, Lecamwasam2024-tn}. In contrast, entanglement has been used to obtain precision enhancement in atomic clocks \cite{Huang_2024}. Additionally, in particular for CV quantum systems, non-Gaussian states have been employed to achieve the HL for phase estimation protocols \cite{Deng2024-uj, PhysRevLett.134.180801}. 

On the other hand, among the different ways of exploiting quantum effects to surpass the performance of a given protocol relative to its classical counterpart, the use of quantum measurements has become an important strategy. Projective measurements have been considered in the design of quantum thermal machines \cite{PhysRevA.104.062210, PhysRevE.110.L052102, PhysRevLett.126.120605, PhysRevLett.120.260601} as well as for proposals of quantum batteries \cite{PhysRevResearch.7.013151, PhysRevA.109.042424}. In each of these applications, fine-tuning the measurement parameters was sufficient to improve their respective figure of merit. Weak measurements, in turn, have been applied in many quantum protocols, ranging from the recent study of quantum realism \cite{Lustosa2025-mr} to their implications in quantum thermodynamics \cite{PhysRevLett.116.080403, Ballesteros_Ferraz2024-mk}. However, quantum measurements are not restricted to being projective or weak; they can be externally adjusted to range from very weak to completely projective. In this case, they are denoted by generalized measurements~\cite{PhysRevE.96.022108, PhysRevA.106.022436,PhysRevA.107.012423}. Applications of generalized measurements include, for instance, quantum cyclic thermal processes \cite{PhysRevA.106.022436,PhysRevA.107.012423} and quantum walks \cite{PhysRevLett.131.150803}. 

In the present work, we consider a single photonic mode as the probe system and explore noncommutative Gaussian measurements during probe-state preparation to investigate the parameter estimation of single-mode Gaussian channels. The probe is initially assumed to be in a thermal state, and the preparation step involves sequential measurements of the position and momentum operators, respectively. We further distinguish between symmetric and asymmetric Gaussian measurements in the preparation protocol state, depending on the measurement uncertainties. The probe then interacts with a single-mode Gaussian channel, and the parameter to be estimated is encoded in the probe's covariance matrix. The QFI is then computed to determine how the Gaussian measurements can be adjusted to improve the estimation. 

The remainder of the paper is organized as follows. Section \ref{sec01} presents the basics of Gaussian quantum mechanics as well as the useful tools for the work. In section \ref{sec02}, we first provide a general discussion of how Gaussian measurements affect the probe during the probe state preparation. Then, we consider two paradigmatic and relevant Gaussian channels, the quantum attenuator and the quantum amplifier. The intrinsic relation between these measurements and their production of coherence in the energy basis of the probe state is also discussed. By introducing a dimensionless parameter that accounts for the asymmetry between the measurements, we directly relate the QFI with the parameter to be estimated to the squeezing in the probe.  We present our conclusions and final remarks in section~\ref{sec03}.

\section{Gaussian quantum framework}\label{sec01}

\textit{States and channels}. We start this section by briefly reviewing the main properties of Gaussian quantum mechanics for a single bosonic mode. This system can be described by the quadrature operator vector $\vec{r}= \left(q,p\right)^T$, with $q$ and $p$ standing for the position and the momentum operators, respectively. Similarly, we can write the quadrature operators in terms of field operators $a\left(a^\dagger\right)$, with $q=\sqrt{\hbar/2m\omega}\left(a^\dagger + a\right)$ and $p = i\sqrt{m\hbar\omega/2}\left(a^\dagger- a\right)$. The quadrature operators satisfy the commutation relation $\left[r_i,r_j\right] = i\hbar\Omega_{ij}$, with $\Omega=i\sigma_{y}$ and $\sigma_y$ is the Pauli matrix.

For a bosonic system, a Gaussian state $\rho$ can be equally described by the characteristic function, defined as
\begin{equation}    
\chi\left(\vec{r}\right)=\exp\left[-\frac{1}{4}\vec{r}^{T}\Omega^{T}\Sigma\Omega\vec{r}+i\vec{r}^{T}\Omega \vec{d}\right],
\end{equation}
with $\vec{d} = \langle \vec{r} \rangle_\rho = \text{Tr}\left[\vec{r}\rho \right]$ and 
\begin{equation}
    \Sigma = \text{Tr}\left[ \left\{ \left(\vec{r} - \vec{d} \right), \left(\vec{r} - \vec{d} \right)^T\right\} \rho\right],
\end{equation}
standing for the statistical mean vector (namely, first moments or displacement vector) and covariance matrix (second moments), respectively, and $\lbrace A,B\rbrace = A\,B + B\,A$ denotes the anticommutator. All information concerning the Gaussian state can be extracted by the statistical moments $\vec{d}$ and $\Sigma$.

Additionally, Gaussian states can evolve in time by the action of quantum channels. In this direction, Gaussian quantum channels (GQCs) represent important physical processes such as loss, squeezing, and interaction with a Markovian environment \cite{serafini2017quantum, eisert2005gaussian}. GQCs are generated by at most second-order Hamiltonians in the quadrature operators \cite{serafini2017quantum}

\begin{equation}
\frac{1}{2}\vec{r}^T\mathrm{H}\vec{r} + \vec{r}^T\vec{r},
\end{equation}
with $\mathrm{H}$ the Hamiltonian symmetric matrix. This is a suitable form for expressing the linear and quadratic contributions to the dynamics. In addition to its theoretical relevance, the second-order Hamiltonian also describes various experimental setups, such as trapped ions, optomechanical systems, and nanomechanical oscillators \cite{serafini2017quantum}. 

Any QGC is characterized by transforming Gaussian states into Gaussian states, with the action on a quantum state $\rho\left(\vec{d}, \Sigma\right)$ described in terms of the statistical moments as
\begin{gather}
\vec{d}\rightarrow\mathcal{M}\vec{d},\\
\Sigma\rightarrow\mathcal{M}\Sigma\mathcal{M}^{T}+\mathcal{N},
\label{generalchannel}
\end{gather}
where $\mathcal{M}$ and $\mathcal{N}$ are $2n\times2n$ real matrices obeying the complete positivity condition
\begin{equation}
\mathcal{N}+i\Omega-i\mathcal{M}\Omega\mathcal{M}^{T}\geq0.
\end{equation}
For $\mathcal{N}=0$, the channel describes a unitary Gaussian operation~\cite{RevModPhys.84.621,serafini2017quantum}, such that single-mode squeezing, displacement, and rotation operations, whereas for $\mathcal{N}\neq0$ they represent attenuation and amplification channels \cite{serafini2017quantum}.

\textit{Generalized measurements}. In recent years, different measurement protocols have attracted considerable interest in quantum information processing and thermodynamics \cite{PhysRevA.106.022436,PhysRevA.107.012423, PhysRevLett.131.150803, PhysRevE.96.022108}.  Consider the measurement of an arbitrary observable $\mathcal{A}$ with eigenvalues $a_\alpha$ as possible results. Generalized measurements are characterized by Hermitian measurement operators $\mathcal{M}_\alpha = \mathcal{M}_\alpha^\dagger$ that satisfy the condition $\sum_\alpha \mathcal{M}_\alpha^2 = \mathbb{I}$ \cite{PhysRevE.96.022108}. For an initial state $\rho_0\left(\lambda\right) = \sum_n p_n^{\text{eq}}\left(\lambda\right)|n,\omega\rangle \langle n,\lambda|$, with $p_n^{\text{eq}}\left(\lambda\right)$ the population of each energy levels and with $\sum_n p_n^{\text{eq}}\left(\lambda \right) = 1$, the post-measurement state is given by
\begin{equation}
    \rho^M\left(\lambda\right) = \sum_\alpha M_\alpha \rho_0\left(\lambda\right) M_\alpha.
    \label{pmstate}
\end{equation}

Here, $\lambda$ corresponds to an arbitrary parameter of the system that can be controlled externally, identified with the single-mode bosonic frequency in the present case. Furthermore, projective measurements are recovered if $\mathcal{M}_\alpha$ agrees with the projection operators onto the eigenspaces of the observable $\mathcal{A}$, but more general measurement schemes can be represented \cite{PhysRevE.96.022108}. We are interested in Gaussian generalized measurements, i.e., those that transform Gaussian initial states into Gaussian final states~\cite{serafini2017quantum}. Gaussian measurements in the position or momentum can be introduced as follows \cite{PhysRevE.96.022108, Cenni2022thermometryof}
\begin{equation}
    M_{\alpha_{r_i}}^{\sigma_{r_i}} = \frac{1}{\left(2\pi \sigma_{r_i}^2  \right)^{1/4}}\exp\left[-\frac{\left(r_i - \alpha_{r_i} \right)^2}{4\sigma_{{r_i}}^2} \right],
    \label{Gaussianmea}
\end{equation}
where $i=[1,2]$ and $r_1 = q$ and $r_2 = p$, standing for the position and momentum Gaussian measurement, respectively, $\sigma_{r_i}$ is the uncertainty for each measurement, and $\alpha_{r_i}$ is the measured position (momentum) eigenvalue. In the particular case where $\sigma_{r_i} \rightarrow 0$, we have a projective measurement, whereas $\sigma_{r_i} \rightarrow \infty$ means the trivial case in which the measurement does nothing to the state. It is important to stress that if the measurement operators $M_{\alpha_{r_i}}$ commute with the Hamiltonian of the system, then $\rho^M\left(\lambda\right)  = \rho_0$. For a single quantum harmonic oscillator, this is not the case for the Gaussian measurements considered here.

\textit{Quantum Fisher Information}. Consider an arbitrary parameter $\theta$, with the squared sensitivity denoted by $\left(\delta \theta \right)^2$. To implement the estimation of $\theta$, we collect $\mathcal{N}$ measurement results $a_i$ of some observable $\mathcal{A}$. The estimation is defined as the variance of the deviation from the true value of $\theta$ of an estimator of $\theta$,  $\theta_\text{est}\left(a_1,...,a_\mathcal{N}\right)$, that depends solely on the measurement results as follows: $\delta \theta^2 = \langle \left[\theta_\text{est}\left(a_1,...,a_\mathcal{N}\right)- \theta\right]^2\rangle_s$, with the notation $\langle...\rangle_{s}$ being the statistical mean. The precision in estimating $\theta$, i.e., $\left(\delta \theta^2\right)$, is bounded from below by the inverse of the quantum Fisher information (QFI),
\begin{equation}
\left(\delta \theta^2\right) \geq \frac{1}{\mathcal{N} \mathcal{I}(\rho_{\theta})},
\end{equation}
where $\mathcal{I}$ denotes the quantum Fisher information for single-parameter estimation~\cite{Huang_2024,Liu_2020,PARIS_09}. For an unbiased estimator, it can be saturated for a large number of measurements and then represents the best reachable bound of sensitivity~\cite{Giovannetti_nat18,Giovannetti_prl21,serafini2017quantum}. The QFI can be written using different distance quantifiers ~\cite{PhysRevA.88.040102,Fuentes2015}. For the Bures distance between two close states $\rho_\theta$ and $\rho_{\theta + \tau}$, defined as 
\begin{equation}
d_{\text{Bures}}\left(\rho_\theta,\rho_{\theta+\tau}\right) = \sqrt{2} \sqrt{1 - \sqrt{\mathcal{F}\left(\rho_\theta, \rho_{\theta + \tau}\right)}},
\end{equation}
with $\tau \ll 1$, the QFI becomes 
\begin{equation}
\mathcal{I}\left(\rho_\theta\right) =\left. 4\left(\frac{\partial d_{\text{Bures}}\left(\rho_\theta,\rho_{\theta+\tau}\right)}{\partial \tau}\right|_{\tau = 0} \right)^2.
    \label{FisherBures}
\end{equation}
The quantity $\mathcal{F}(\rho_{\theta},\rho_{\sigma})=\left(\text{Tr}\sqrt{\sqrt{\rho_{\theta}}\rho_{\sigma}\sqrt{\rho_{\theta}}}\right)^{2}$ is the fidelity between the two states mentioned. For Gaussian states, the fidelity can be completely written only in terms of the first moments and covariance matrix of the states as follows~\cite{RevModPhys.84.621}
\begin{equation}
\mathcal{F}\left(\rho_\theta,\rho_{\sigma}\right) = \frac{2}{\sqrt{\Delta + \delta} - \sqrt{\delta}} \exp\left[-\frac{1}{2} \Delta\vec{d}^T\left(\Sigma_\theta + \Sigma_{\sigma}\right)^{-1} \Delta\vec{d} \right],
\end{equation}
with 
\begin{equation}
\Delta \equiv \det\left[\Sigma_\theta + \Sigma_{\sigma} \right], 
\end{equation}
\begin{equation}
\delta \equiv (\text{det}\Sigma_{\theta}-1)(\text{det}\Sigma_{\sigma}-1),
\end{equation}
and 
\begin{equation}
\Delta\vec{d} = \vec{d}_\theta - \vec{d}_{\sigma}.
\end{equation}

Similarly, the QFI can also be written in terms of fidelity \cite{PhysRevA.88.040102}. By restricting to a single-mode Gaussian state and considering the expansion of the fidelity up to the second order in $\tau$, the QFI is finally written as
\begin{equation}
\mathcal{I}\left(\rho_{\theta}\right)=\frac{1}{2}\frac{\text{Tr}\left[\left(\Sigma_{\theta}^{-1}\Sigma_{\theta}'\right)^{2}\right]}{1+P_{\theta}^{2}}+2\frac{(P_{\theta}^{'})^{2}}{1-P_{\theta}^{4}}+\Delta\vec{d}'^{T}\Sigma_{\theta}^{-1}\Delta\vec{d}',
     \label{Fisher}
\end{equation}
where $P_\theta = |\Sigma_\theta|^{-2}$ represents the purity of the single-mode Gaussian state, $\Sigma_{\theta}^{-1}$  is the inverse matrix of $\Sigma_\theta$, $\Sigma_\theta'$ and $P_{\theta}^{'}$ denotes the differentiation of the covariance matrix and purity concerning the parameter $\theta$, respectively. From Eqs. (\ref{FisherBures}) and (\ref{Fisher}), we conclude that the more sensitive the probe is to small deviations in the arbitrary parameter $\theta$, the higher the precision in the estimation. In addition, Eq.(\ref{Fisher}) shows that the QFI depends on three terms. The first term describes how the covariance matrix dynamically depends on the encoding parameter $\theta$. The second term represents the dynamics of purity as $\theta$ varies. The third term accounts for the contribution of the first-moment dynamics of the Gaussian state to the estimated parameter.

\section{Effects of Gaussian measurements on the QFI}\label{sec02}

This section investigates how Gaussian measurements used in the probe-state preparation protocol affect the estimation of Gaussian channel parameters. We consider a photonic system initially represented by a single-mode Gaussian state $\rho_{\text{th}}\left(\vec{d}_0, \Sigma_0\right)$, with $\vec{d}_0$ and $\Sigma_0$ the initial first moments and covariance matrix, respectively. To do so, we assume the following protocol illustrated by the Fig. \ref{Scheme}:
\begin{figure}[h]
\includegraphics[width=1.0\linewidth]{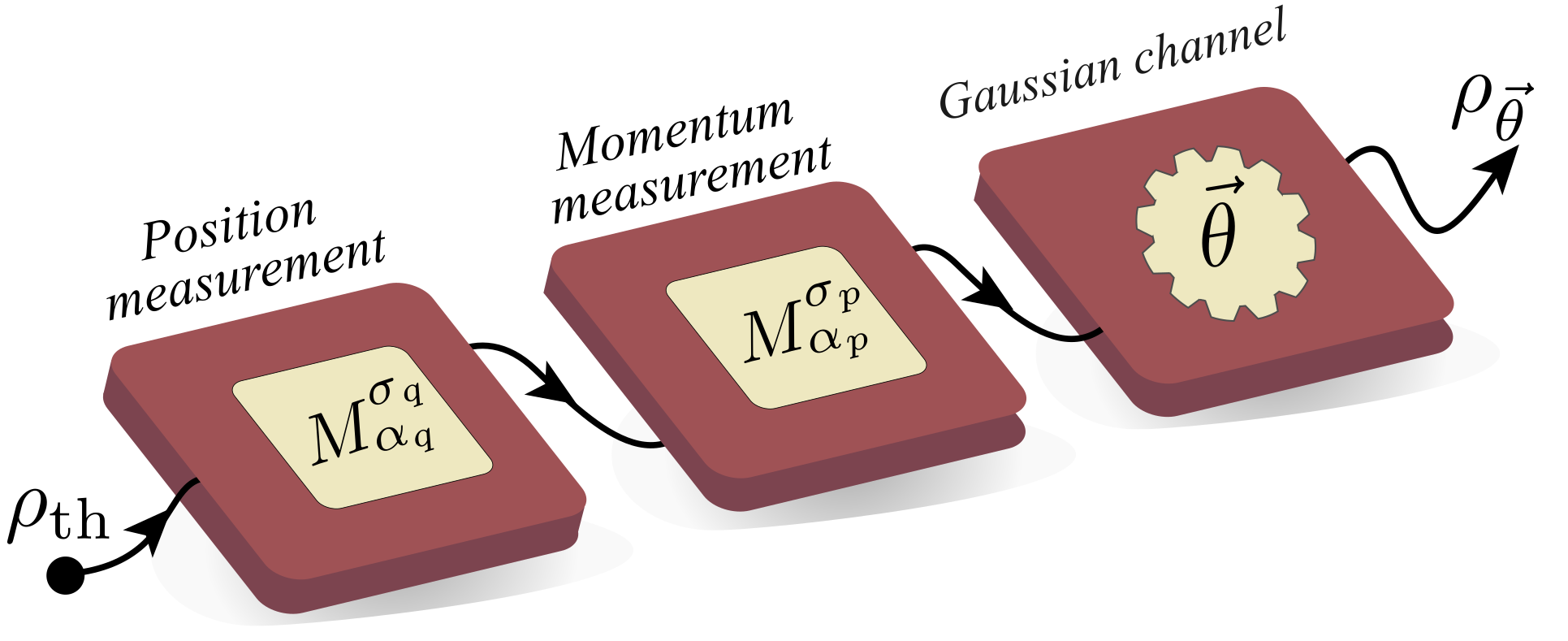}  
\caption{Illustration of the proposed probe state protocol. The single-mode Gaussian state is initially in a thermal state with inverse temperature $\beta$. The first Gaussian measurement is performed in the position observable with an uncertainty $\sigma_q$, followed by a second Gaussian measurement in the momentum observable, with an uncertainty $\sigma_p$. Then, the probe interacts with a single-mode Gaussian channel with a parameter vector $\vec{\theta}$, and we use the final probe state to interrogate a specific channel parameter. }
\label{Scheme}
\end{figure}

\textit{1. State preparation:} Starting with a single photonic mode prepared in a thermal state, $\rho_{\text{th}} = \exp\left[-\beta H \right]/Z$, with $H = \hbar \omega a^\dagger a$ the photonic Hamiltonian with frequency $\omega$ and $Z = \text{Tr}\lbrace\exp\left[-\beta H \right]\rbrace$ the partition function, the probe state preparation protocol involves the use of two noncommutative Gaussian measurements, in the observables position and momentum, respectively denoted by $M_{\alpha_{q}}^{\sigma_q}$ and $M_{\alpha_{p}}^{\sigma_p}$, which in turn are given by Eq. (\ref{Gaussianmea}), with $\sigma_q$ and $\sigma_p$ the respective uncertainty for each measurement. After this protocol, the probe state becomes 
\begin{equation}
\rho_{\text{probe}} =M_{\alpha_{p}}^{\sigma_p} M_{\alpha_{q}}^{\sigma_q}\rho_{\text{th}}\left(M_{\alpha_{q}}^{\sigma_q}\right)^\dagger\left(M_{\alpha_{p}}^{\sigma_p}\right)^\dagger.
    \label{probe}
\end{equation}

Although it is mathematically possible to obtain the state $\rho_{\text{probe}}$ analytically, we observe that the probe state remains Gaussian during all the time of the estimation protocol, and then it is also completely characterized by its first moments and the covariance matrix, $\rho_{\text{probe}} = \rho_{\text{probe}}\left(\vec{d}_{\vec{\theta}}, \Sigma_{\vec{\theta}}\right)$. Besides, the QFI for a Gaussian single mode depends solely on $\vec{d}_{\vec{\theta}}$ and $\Sigma_{\vec{\theta}}$. Since the first moments are null for the initial thermal state and a GQM does not modify them, we must obtain $\Sigma_{\vec{\theta}}$. After the probe state preparation protocol, the final covariance matrix elements are given by (See Appendix~\ref{Appendix})
\begin{eqnarray}
    \Sigma_{11}^{\text{probe}} &=&   \frac{\hbar\coth\left(\beta \hbar \omega/2\right)}{m\omega\sigma_{p}^{2}}  ,\nonumber \\
    \Sigma_{12}^{\text{probe}} &=& \Sigma_{21}^{\text{probe}} = 0    ,\nonumber\\
    \Sigma_{22}^{\text{probe}} &=& \frac{\hbar m\coth\left(\beta \hbar \omega/2\right))}{\sigma_{q}^{2}},
\label{aftermeasurement}
\end{eqnarray}
forming the matrix $\Sigma^{\text{probe}}$, with $\sigma_{q}$ and $\sigma_{p}$ playing the role of the measurement resources, which can be appropriately tuned to improve the QFI for a given parameter estimation. It is important to note that, in the present case, we assume that $\sigma_{q}$ and $\sigma_{p}$ are completely known. Specifically, it is supposed that the uncertainties are experimentally controlled, possibly using ancillary systems to implement them \cite{PhysRevLett.108.253601}.

\textit{2. Acquiring information of a Gaussian channel:} Let us assume a GQC which is characterized by the vector of parameters $\vec{\theta} = \left(\theta_1,...,\theta_Q\right)$, with $Q$ the number of channel parameters. From Eq. (\ref{generalchannel}), after the photonic mode has passed through a QGC, the final state is described by the statistical moments
\begin{gather}
\vec{d}_{\vec{\theta}} = {M}_{\vec{\theta}}\vec{d}^{\text{probe}},\nonumber\\
\Sigma_{\vec{\theta}} = \mathcal{M}_{\vec{\theta}}\Sigma^{\text{probe}}\mathcal{M}_{\vec{\theta}}^{T}+\mathcal{N}_{\vec{\theta}},
\label{finalstate}
\end{gather}
with the matrices $\mathcal{M}_{\vec{\theta}}$ and $\mathcal{N}_{\vec{\theta}}$ dictating the final state structure and carrying all information of the parameter to be estimated. By using Eqs (\ref{aftermeasurement}) in Eq. (\ref{finalstate}), all information regarding the estimation of the parameter $\theta_i$ is contained in the covariance matrix $\Sigma_{\vec{\theta}}= \Sigma_{\vec{\theta}}\left(\sigma_q,\sigma_p\right)$, since $\vec{d}_0 = 0$ for the initial thermal state, $\rho_{\text{th}}$.

\textit{3. Lower bounds for the parameters to be estimated:} This last step covers the calculation of the possible lower bound for the precision of the parameter $\theta_i$ using the quantum Fisher information~(\ref{Fisher}) and the role played by the probe state preparation in this process. 

In the following, we consider two paradigmatic examples of Gaussian channels for a single photonic mode, with wide applications in quantum information processing. We stress that, the parameters involving the probe state preparation, i.e., the vector of parameters $\vec{\theta}^{\text{probe}} = \left(\beta,\omega, \sigma_q, \sigma_p \right)$ are assumed to be completely known, such that we are interested only in estimating the channel parameters.

\subsection*{The quantum attenuator channel (QAttC)}

The first relevant single-mode Gaussian channel that we would like to investigate is the quantum attenuator channel $\mathcal{E}\left(\varphi, \bar{m} \right)$, which represents the interaction of a single-mode $\rho_S$ with an auxiliary mode $\rho_A$ initially prepared in a thermal state with average thermal number $\bar{m}$. Here, the parameter $\varphi$ mediates a beam-splitter interaction, and it can be directly written in terms of a transmissivity coefficient $\eta = \cos^2(\varphi)$. Furthermore, the attenuator channel is often used to simulate losses in a single mode, and in this case, the parameter $\varphi$ is linked to the interaction time between the mode and its environment~\cite{Giovannetti_prl21,serafini2017quantum}. In terms of Eqs. (\ref{finalstate}), the matrices describing the QAC are given by
\begin{eqnarray}
{M}_{\vec{\theta}}^{\text{att}} &=& \cos(\varphi) \mathbb{I}_2,\nonumber\\
{N}_{\vec{\theta}}^{\text{att}} &=& \sin^2(\varphi)\left(2\bar{m} + 1\right)\mathbb{I}_2,
\end{eqnarray}
with the vector of parameters  $\vec{\theta}^{\text{att}} = \left(\varphi, \bar{m} \right)$ for the QAttC. As the first moments of the probe state are zero, the parameter estimation depends solely on the covariance matrix $\Sigma_{\vec{\theta}^{\text{att}}} = \textbf{diag}\left(\nu_1,\nu_2\right)$, with the symplectic eigenvalues 
\begin{eqnarray}
    \nu_1 &=& \frac{\coth\left(\beta \hbar \omega/2\right)\cos^2 (\varphi)}{\sigma_p^2} + \left(2\bar{m}+1\right)\sin^2(\varphi),\nonumber\\
    \nu_2 &=& \frac{\coth\left(\beta \hbar \omega/2\right)\cos^2(\varphi)}{\sigma_q^2} + \left(2\bar{m}+1\right)\sin^2(\varphi).
    \label{eigenvaluesatt}
\end{eqnarray}

For completeness, the QFI for the estimation of $\varphi$ and $\bar{m}$ are given below

\begin{widetext}
\begin{eqnarray}
\mathcal{I}^{\varphi}_{\text{att}}(\varphi,\bar{m}) &=& \frac{\sin^{2}\left(2\varphi\right)}{2}\left[\frac{\nu_{2}^{2}h_{p}^{2}+\nu_{1}^{2}h_{q}^{2}}{\nu_{1}\nu_{2}\left(\nu_{1}\nu_{2}+1\right)}+\frac{\left(h_{q}\nu_{1}+h_{p}\nu_{2}\right)^{2}}{\nu_{1}\nu_{2}\left(\nu_{1}^{2}\nu_{2}^{2}-1\right)}\right],\nonumber\\
\mathcal{I}^{\bar{m}}_{\text{att}}\left(\varphi,\bar{m}\right) &=& 2\sin^{4}(\varphi)\left[\frac{\nu_{1}^{2}+\nu_{2}^{2}}{\nu_{1}\nu_{2}\left(\nu_{1}\nu_{2}+1\right)}+\frac{\left(\nu_{1}+\nu_{2}\right)^{2}}{\nu_{1}^{2}\nu_{2}^{2}-1}\right],
\label{QFIatteq}
\end{eqnarray}
\end{widetext}
where the auxiliary functions $h_p$ and $h_q$ defined as
\begin{eqnarray}
h_p &=& \left(2\bar{m}+1\right)-\frac{\coth\left(\beta\hbar\omega/2\right)}{\sigma_{p}^{2}},\nonumber\\
h_q &=& \left(2\bar{m}+1\right)-\frac{\coth\left(\beta\hbar\omega/2\right)}{\sigma_{q}^{2}}.
\end{eqnarray}

\begin{figure}[!h]
\includegraphics[scale=0.135]{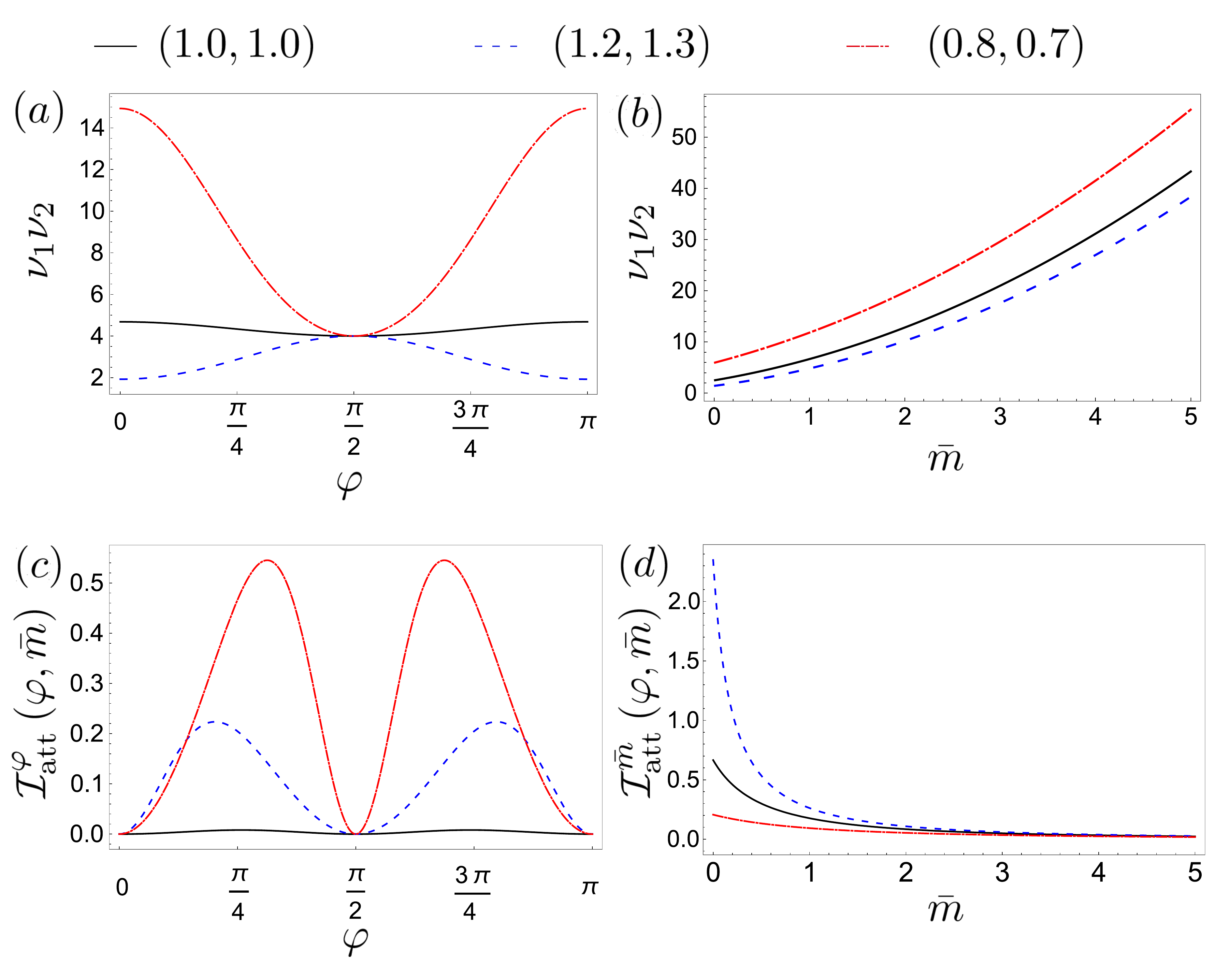}  
\caption{Behavior of the symplectic eigenvalues and quantum Fisher information for the QAttC. (a) and (b) illustrate the behavior of the product between the symplectic eigenvalues, $\nu_1\nu_2$, as a function of $\varphi$ and $\bar{m}$, respectively. (c) shows the QFI for the estimation of $\varphi$, with a fixed value of $\bar{m} = 0.5$, while (d) shows the QFI for the estimation of $\bar{m}$, with a fixed value of $\varphi = \pi/4$. We set $\beta = \omega = 1$, and the values of $\sigma_q$ and $\sigma_p$ such that the product $\sigma_q\sigma_p$ satisfies the uncertainty principle for the probe state.}
\label{QFI_att01}
\end{figure}

Figure \ref{QFI_att01} depicts the QFI for the quantum attenuator channel for the estimation of $\varphi$ and $\bar{m}$. The behavior of the symplectic eigenvalues $\nu_1\nu_2$ as a function of $\varphi$ is illustrated in Fig. \ref{QFI_att01}-(a), and it is direct to conclude that $\mathcal{I}^{\varphi}_{\text{att}}(\varphi,\bar{m})$ depends on the shape of $\nu_1\nu_2$. In principle, it could be expected from the second term for $\mathcal{I}^{\varphi}_{\text{att}}(\varphi,\bar{m})$ that it would diverge when $\nu_1\nu_2 \rightarrow 1$. However, it is important to stress that this term is modulated by the factor $\cos(\varphi) \sin(\varphi)$. In Fig. \ref{QFI_att01}-(c), it is shown $\mathcal{I}^{\varphi}_{\text{att}}(\varphi,\bar{m})$ as a function of $\varphi$ and for a fixed value of $\bar{m} = 0.5$. By comparing the dashed blue and dashed-dotted red lines with the solid black line (the purely thermal case), we observe that the probe-state preparation using two non-commutative measurements can indeed improve the QFI. Here, we consider an appropriate parametrization in the uncertainty $\sigma_{r_i}$, such that the condition $\sigma_{r_i} \rightarrow 1$ represents the scenario in which the measurements do nothing on the initial probe state (See Appendix~\ref{Appendix}). On the other hand, Fig. \ref{QFI_att01}-(b) illustrates the behavior of the symplectic eigenvalues $\nu_1\nu_2$ as a function of $\bar{m}$, for a fixed value of $\varphi = \pi/4$. In particular for $\varphi = \pi/2$ the quantity $\nu_1\nu_2 \rightarrow 1$ as $\bar{m} \rightarrow 0$.

Fig. \ref{QFI_att01}-(d) shows $\mathcal{I}^{\bar{m}}_{\text{att}}\left(\varphi,\bar{m}\right)$ as a function of $\bar{m}$ and for a fixed $\varphi = \pi/4$. It is possible to observe that, in contrast to the estimation of $\varphi$, here it is not true that $\sigma_q\neq 1$ and $\sigma_p\neq 1$ will always represent an advantage in terms of the QFI, mainly because we are quantifying the lower bound of the channel temperature, which is more sensitive to the probe temperature. Note that from the second term in $\mathcal{I}^{\bar{m}}_{\text{att}}\left(\varphi,\bar{m}\right)$, Eq. (\ref{QFIatteq}), the QFI will enhance as the quantity $\nu_1\nu_2$  approaches 1. In particular, for $\varphi = \pi/4$, $\nu_1\nu_2 \rightarrow 1$ as $\bar{m}$ approaches zero, leading to $\mathcal{I}^{\bar{m}}_{\text{att}}\left(\varphi,\bar{m}\right) \rightarrow \infty$, in agreement with previous results \cite{serafini2017quantum}. Furthermore, setting $\varphi = \pi/2$ is of particular interest since it corresponds to a complete swap between the probe and environment states.

\subsection*{The quantum amplifier channel (QAmpC)}

The second single-mode Gaussian channel, whose characterization could benefit from the state-preparation protocol with non-commutative Gaussian measurements, is the so-called quantum amplifier channel (QAmpC). The relevance of this quantum channel is that it represents the introduction of noise into the initial state, enhancing its amplitude \cite{RevModPhys.84.621, serafini2017quantum}. An interesting application of the QAmpC is in a Gaussian-cloning machine model, where it is employed in conjunction with a balanced beam splitter \cite{RevModPhys.84.621}. The QAmpC is directly connected to the two-mode squeezing operation, $S_2\left(r\right) = \exp\left[r_g \left(a b - a^\dagger b^\dagger\right) \right]$, with squeezing parameter $r_g$, which is parametrically written in terms of the amplification channel gain coefficient $g \geq 1$ as $g = \cosh^2 (r_g)$.  In analogy to the attenuator channel, here we can also represent the action over the first moments and the covariance matrix using the matrices
\begin{eqnarray}
    {M}_{\vec{\theta}}^{\text{amp}} &=& \cosh (r_g) \mathbb{I}_2,\nonumber\\
    {N}_{\vec{\theta}}^{\text{amp}} &=& \sinh^2(r_g)\left(2\bar{m} + 1\right)\mathbb{I}_2,
\end{eqnarray}
with the vector of parameters $\vec{\theta}^{\text{amp}} = \left(r_g,\bar{m}\right)$ for the QAmpC. Again, the first moments are zero, with all information of the probe state after it had passed through the QAmpC, encoded in the covariance matrix $\Sigma_{\vec{\theta}^{\text{amp}}} = \textbf{diag}\left(\mu_1,\mu_2\right)$, where the symplectic eigenvalues are given by
\begin{eqnarray}
    \mu_1 &=& \frac{\coth\left(\beta \hbar \omega/2\right)\cosh^2 (r_g)}{\sigma_p^2} + \left(2\bar{m}+1\right)\sinh^2(r_g),\nonumber\\
    \mu_2 &=& \frac{\coth\left(\beta \hbar \omega/2\right)\cosh^2 (r_g)}{\sigma_q^2} + \left(2\bar{m}+1\right)\sinh^2 (r_g).
    \label{eigenvaluesatt}
\end{eqnarray}

It is also possible to express analytically the QFI for the estimation of $r_g$ and $\bar{m}$ for the QAmpC in terms of the symplectic eigenvalues. The expressions are
\begin{widetext}
\begin{eqnarray}
    \mathcal{I}^{r_{g}}_{\text{amp}} \left(r_g,\bar{m}\right) &=& \frac{\sinh^{2}\left(2r_{g}\right)}{2}\left[\left(\frac{\mu_{2}^{2}f_{p}^{2}+\mu_{1}^{2}f_{q}^{2}}{\mu_{1}\mu_{2}\left(\mu_{1}\mu_{2}+1\right)}\right)+\frac{\left(f_{q}\mu_{1}+f_{p}\mu_{2}\right)^{2}}{\left(\mu_{1}\mu_{2}\left(\mu_{1}^{2}\mu_{2}^{2}-1\right)\right)}\right]    ,\nonumber\\
    \mathcal{I}^{\bar{m}}_{\text{amp}} \left(r_g,\bar{m}\right) &=&  2\sinh^{4} (r_{g})\left(\frac{\left(\mu_{1}^{2}+\mu_{2}^{2}\right)}{\mu_{1}\mu_{2}\left(\mu_{1}\mu_{2}+1\right)}+\frac{\left(\mu_{1}+\mu_{2}\right)^{2}}{\mu_{1}\mu_{2}\left(\mu_{1}^{2}\mu_{2}^{2}-1\right)}\right), 
    \label{QFIampeq}
\end{eqnarray}
\end{widetext}
with the functions $f_p$ and $f_q$ defined by
\begin{eqnarray}
    f_p &=& \left(2\bar{m}+1\right)+\frac{\coth\left(\beta\hbar\omega/2\right)}{\sigma_{p}^{2}},\nonumber\\
    f_q &=& \left(2\bar{m}+1\right)+\frac{\coth\left(\beta\hbar\omega/2\right)}{\sigma_{q}^{2}}.
\end{eqnarray}

\begin{figure}[H]
\centering
\includegraphics[scale=0.135]{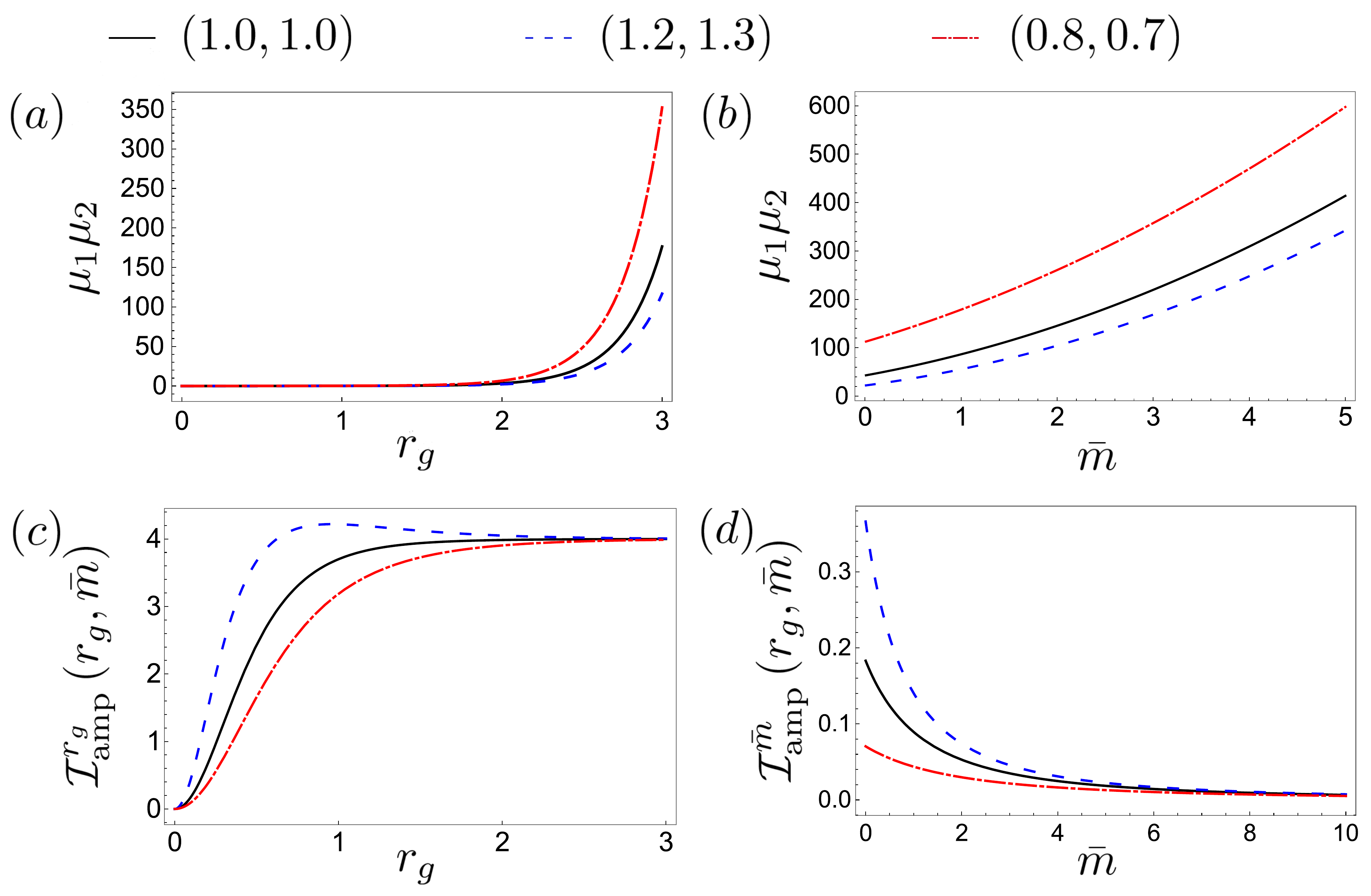}  
\caption{Behavior of the symplectic eigenvalues and quantum Fisher information for the QAmpC. (a) and (b) illustrate the behavior of the product between the symplectic eigenvalues, $\mu_1\mu_2$, as a function of $r_g$ and $\bar{m}$, respectively. (c) shows the QFI for the estimation of $r_g$, with a fixed value of $\bar{m} = 0.5$, while (d) shows the QFI for the estimation of $\bar{m}$, with a fixed value of $r_g = 1.0$. We have considered $\beta = \omega = 1$, and the values of $\sigma_q$ and $\sigma_p$ such that the product $\sigma_q\sigma_p$ satisfies the uncertainty principle for the probe state.}
\label{QFI_amp01}
\end{figure}

Figure \ref{QFI_amp01}  shows the behavior of the product $\mu_1\mu_2$ as a function of the squeezing parameter $r_g$ (Fig. \ref{QFI_amp01}-(a)) and the average thermal number $\bar{m}$ (Fig. \ref{QFI_amp01}-(b)), whereas the QFI for the amplification channel is depicted for $r_g$ (Fig. \ref{QFI_amp01}-(c)) and for $\bar{m}$ (Fig. \ref{QFI_amp01}-(d)). We have chosen values of $\sigma_q$ and $\sigma_p$ corresponding to an effective higher (red dashed-dotted lines) and lower (blue dashed lines) average thermal number. By fixing $\bar{m} = 0.5$ and focusing on the QFI for $r_g$, we observe from Fig. \ref{QFI_amp01}-(c) that Gaussian measurements provide an enhancement in the QFI for an effective lower temperature, indicating that the reduction of thermal fluctuations is beneficial for the estimation of $r_g$. In addition, as the determinant of the covariance matrix diverges as a function of $r_g$, the QFI goes to an asymptotic value irrespective of the choices of $\sigma_q$ and $\sigma_p$. On the other hand, the behavior of the determinant of the covariance matrix as well as the QFI for the estimation of $\bar{m}$ for a fixed value of $r_g = 1.0$ is basically identical to the attenuator channel, as expected. 

\subsection*{Role played by quantum coherence}

We have shown that by fine-tuning the uncertainties associated with non-commutative measurements, namely, $\sigma_q$ and $\sigma_p$, it is possible to increase the QFI for parameters of Gaussian channels, thereby possibly improving the precision of a given parameter. It is then pertinent to investigate in detail how these measurements affect the quantum probe before and after it passes through the Gaussian channel. Before the system interacts with the channel, the probe is completely described by the covariance matrix in Eq. (\ref{aftermeasurement}), such that the probe purity is given by
\begin{equation}
\mathcal{P}\left(\rho^{\text{probe}}\right) = \frac{\sigma_q \sigma_p}{\left(2\bar{n}+1\right)}.
\label{purity}
\end{equation}

Note that if $\bar{n} = 0$ and $\sigma_q = \sigma_p = 1$, which means that the probe is initially in the ground state and no measurements are applied on the system during the preparation protocol, $\mathcal{P}\left(\rho^{\text{probe}}\right) = 1$. On the other hand, increasing the temperature increases $\bar{n}$, thereby reducing purity, as expected. However, Eq. (\ref{purity}) shows another way to control the purity, namely by adjusting the measurement uncertainties $\sigma_q$ and $\sigma_p$. Adjusting $\sigma_q$ and $\sigma_p$ then corresponds to different effective average thermal numbers or, equivalently, to different effective temperatures. The concept of effective temperature is often associated with the activation of non-diagonal elements of the density operator during a given unitary evolution, thereby enabling transitions between different energy eigenstates. These non-diagonal elements are the core of quantum coherence, a key quantum resource widely employed in different protocols \cite{Streltsov2017}. The operational meaning of quantum coherence was first introduced in Ref. \cite{Baumgratz2014}, where it is quantified using the $\ell_1$-norm or the relative entropy of coherence. For continuous variable (CV) systems, the $\ell_1$-norm may be tricky, and the relative entropy of coherence is more convenient, in particular for Gaussian states. By choosing the coherence in the eigenstates of energy, i.e., the Fock basis, a general single-mode state is written as
\begin{equation}
\rho = \sum_{i,j = 0}^\infty\rho_{ij}|i\rangle\langle j|,
\end{equation}
where $\sum_{i=0}^\infty \rho_{i,i} = 1$ ensures a physical state. 

The relative entropy quantifies the coherence of a CV state $\rho$ by comparing it with a diagonal reference state $\rho^{\textbf{ref}}$ at the same basis and is given by \cite{Baumgratz2014}
\begin{equation}
    \mathcal{C}\left(\rho\right) = S\left(\rho^{\textbf{ref}}\right) - S\left(\rho \right), 
\end{equation}
with $S\left( \cdot \right)$ the von Neumann entropy. In the present case, the state $\rho$ corresponds to the probe state, which is a single-mode Gaussian state with the parameter information printed into the first and second statistical moments, $\rho = \rho_{\text{probe}}\left(\vec{d}_{\vec{\theta}}, \Sigma_{\vec{\theta}}\right)$. In such a case, the von Neumann entropy is given by
\begin{equation}
    S\left(\rho\right) = \frac{\nu+1}{2}\log\left(\frac{\nu+1}{2}\right) - \frac{\nu-1}{2}\log\left(\frac{\nu-1}{2}\right),
\end{equation}
with $\nu = \sqrt{\det \Sigma_{\vec{\theta}}}$ the corresponding eigenvalues of $\Sigma_{\vec{\theta}}$. In addition, Ref. \cite{Xu2016} has shown that for a single-mode Gaussian state, the diagonal reference state $\rho^{\textbf{ref}}$ is a thermal state, with average thermal number $\bar{N}({\rho^{\textbf{ref}}}) = \left[\text{Tr}\left(\Sigma_{\vec{\theta}}\right) + d^2_{1,\theta} + d^2_{2,\theta} - 2 \right]/4$.

Given the estimation of an arbitrary parameter $\theta$ employing a probe state which uses quantum coherence as a resource to the task, the connection between the QFI and the change of coherence in the parameter space was discussed in Ref. \cite{Santos2025}. As can be observed from Eqs. (\ref{aftermeasurement}), The use of position and momentum Gaussian measurements in the preparation protocol can introduce squeezing into the probe, thereby representing coherence in the energy basis by modifying the covariance matrix. This can be denoted by $\Sigma_{\vec{\theta}} = \Sigma_{\vec{\theta}}\left[\mathcal{C\left(\rho_{\text{probe}}\right)}\right]$. Formally, we can write
\begin{equation}
    \partial_\theta \Sigma_{\vec{\theta}} = \frac{\partial \Sigma_{\vec{\theta}}}{\partial \mathcal{C}}\frac{\partial \mathcal{C}}{\partial \theta},
    \label{coherenceder}
\end{equation}
indicating that the response of the covariance matrix due to a parameter change depends on the respective response of the coherence. Thus, by comparing Eqs. (\ref{Fisher}) and (\ref{coherenceder}), we see that if the coherence of the probe state does not change with the parameter to be estimated, then the QFI depends only on the thermal resources of the initial probe state. 

\begin{figure}[H]
\centering
\includegraphics[width=0.8\linewidth]{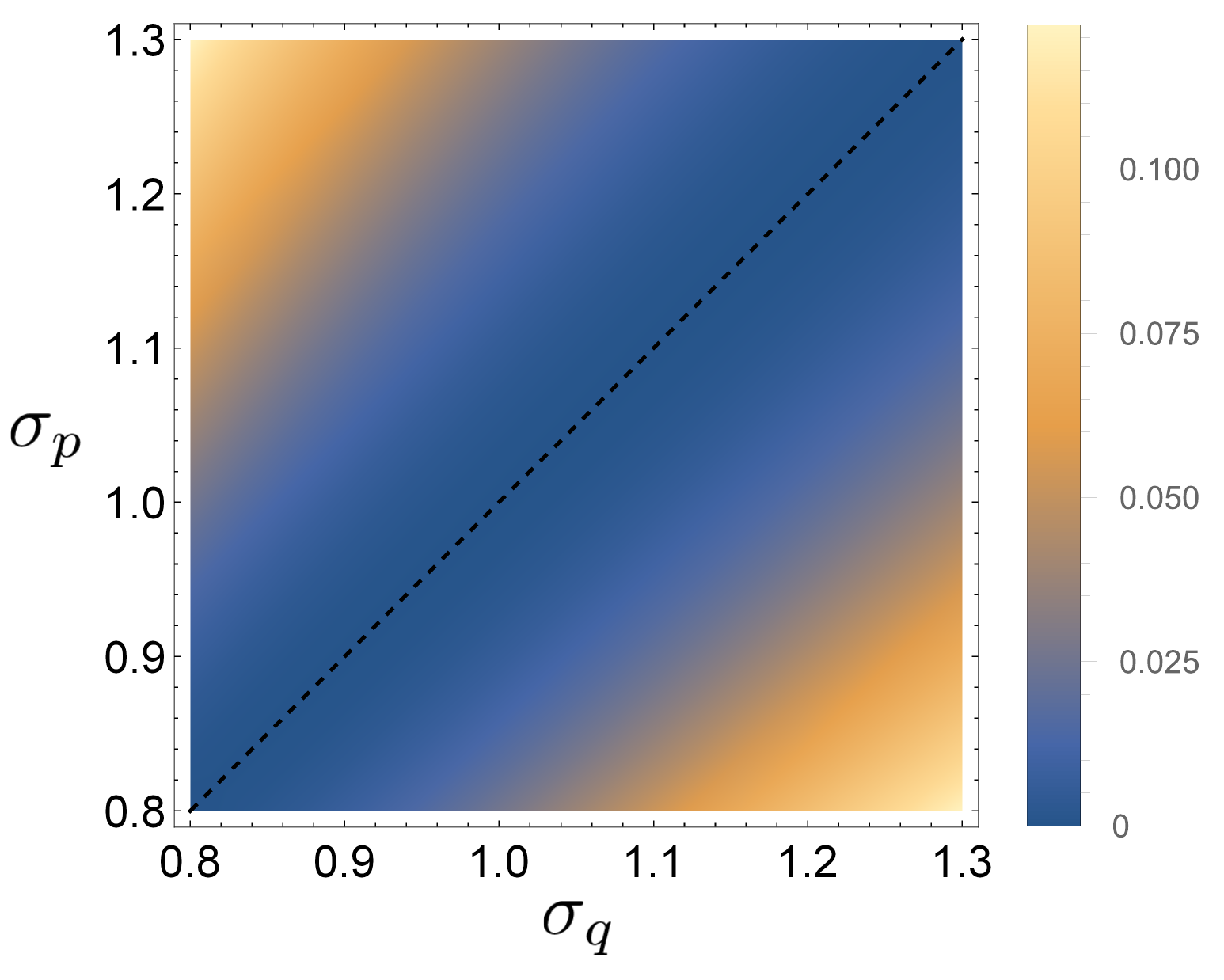}  
\caption{Quantum coherence of the probe state as a function of the uncertainties $\sigma_q$ and $\sigma_p$ after the two Gaussian measurements in the probe state preparation and before any quantum channel. The dashed black straight line represents the symmetric scenario in which $\sigma_q = \sigma_p$, with no coherence being produced. The probe initial parameters are the same as in Fig. \ref{QFI_att01} and Fig. \ref{QFI_amp01}.}
\label{cohein}
\end{figure}

To understand the role played by the quantum coherence in the parameter estimation of Gaussian channels, we first show in Fig. \ref{cohein} a density plot of the quantum coherence for the probe state after the two Gaussian measurements and before any channel, as a function of $\sigma_q$ and $\sigma_p$. The symmetric scenario ($\sigma_q = \sigma_p$) is indicated by the black dashed line and does not represent a real boost from the measurements, since it can be achieved by simply adjusting the probe's initial temperature. On the other hand, the asymmetric scenario ($\sigma_q \neq \sigma_p$) is what we designate by a true boost due to the measurements, since it is directly associated with the production of quantum coherence.

\begin{figure}[H]
\centering
\includegraphics[width=1.0\linewidth]{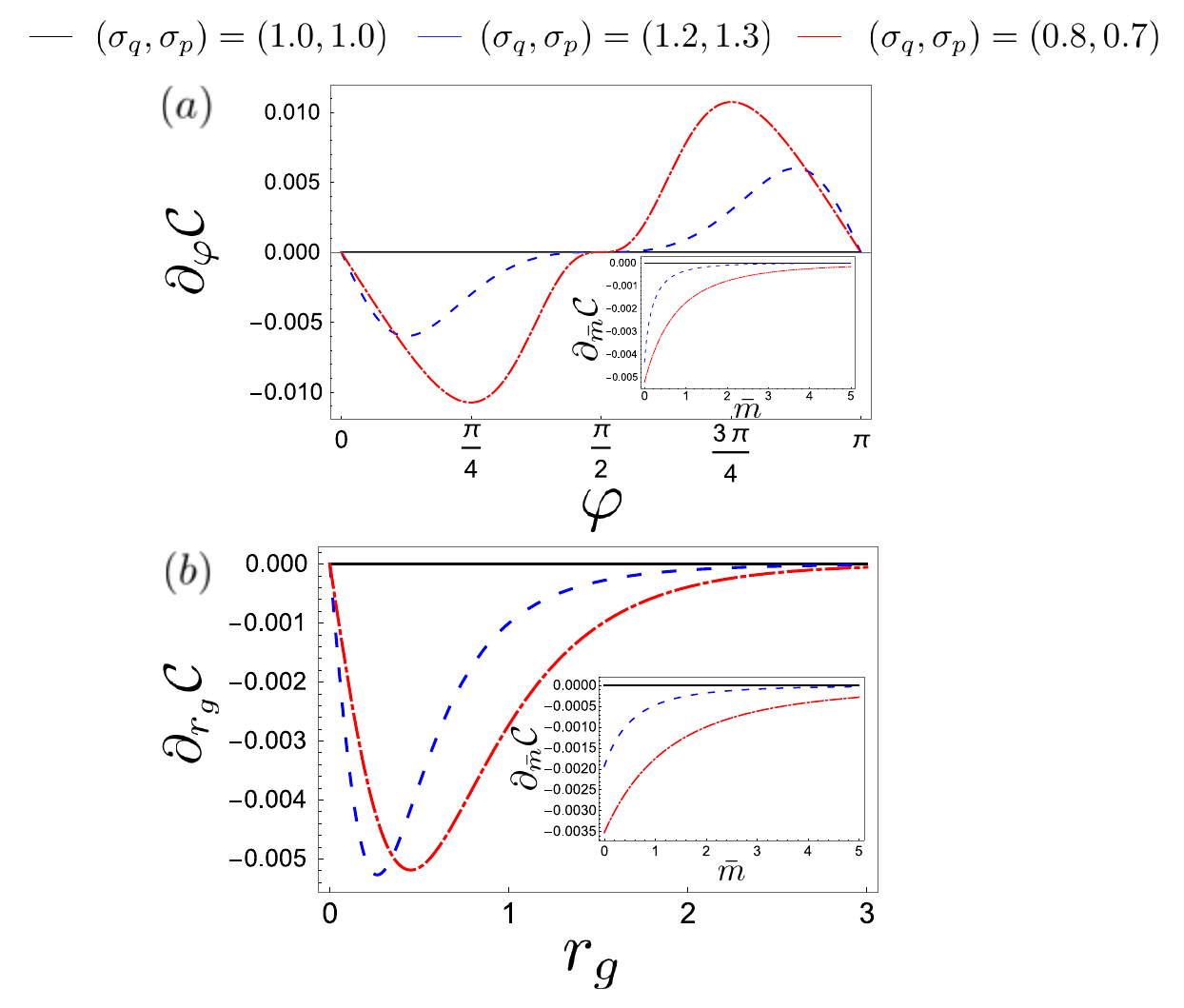}
\caption{Derivative of the quantum coherence with respect to different channel parameters. (a) Derivative of the coherence as a function of the parameter $\varphi$ for the attenuator channel, with the inset the corresponding derivative for the channel average thermal number $\bar{m}$. (b) Derivative of the coherence as a function of the parameter $r_g$ for the amplifier channel, with the inset the corresponding derivative channel average thermal number $\bar{m}$. The probe initial parameters are the same as in Fig. \ref{QFI_att01} Fig. \ref{QFI_amp01}.}
\label{CoheDer}
\end{figure}

Now, we focus on the behavior of the quantum coherence as the probe interacts with the Gaussian channels. More specifically, we are interested in the derivative of the coherence with respect to the different parameters for which we have computed the QFI. Figure \ref{CoheDer} depicts the derivative of quantum coherence as a function of different channel parameters. For the estimation of $\varphi$ in the attenuator channel, the more sensitive the quantum coherence is relative to the beam-splitter angle, the higher the QFI for $\varphi$. In contrast, this is not the role observed for the estimation of $r_g$ in the amplifier channel. The derivative of the quantum coherence still serves as a witness for the QFI, in the sense that as it goes to zero for large $r_g$, $\mathcal{I}_{\text{amp}}\left(r_g\right)$ approaches a constant value. However, the difference between the QFI for the case of $\sigma_q = 1.2$ (blue dashed line) and $\sigma_q = 0.8$ shows that there exists a limit of quantum coherence to be introduced during the preparation protocol above which the QFI becomes worse.

Finally, the insets in Figs. \ref{CoheDer}-(a) and \ref{CoheDer}-(b) illustrate the derivative of the quantum coherence with respect to the average thermal number $\bar{m}$, for the attenuator and amplifier channel, respectively. Both behaviors are similar and indicate that the increase of $\bar{m}$ only reduces the quantum coherence, as expected, and that the derivative goes to zero asymptotically. 

\begin{figure}[H]
\centering
\includegraphics[scale=0.25]{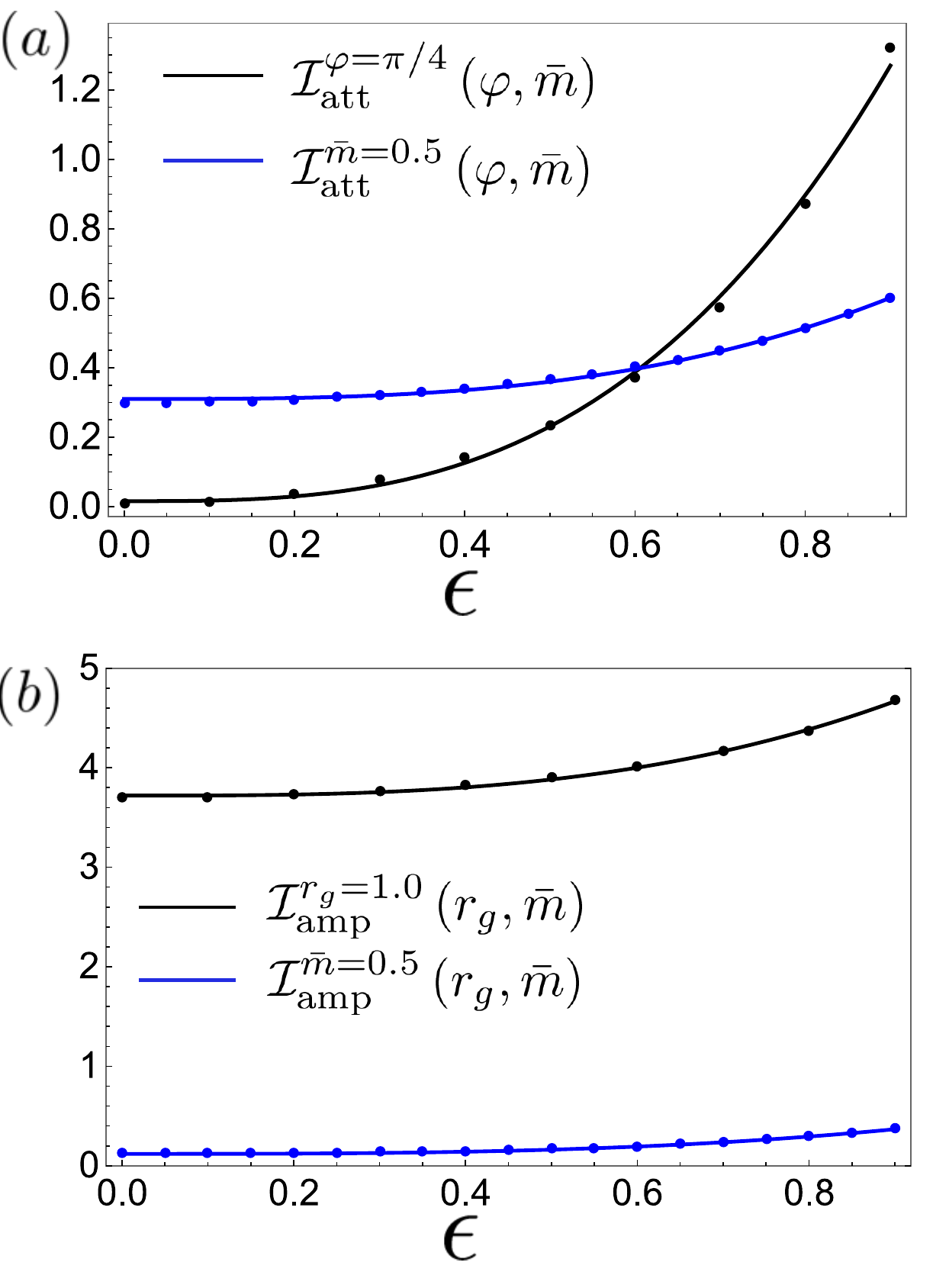}  
\caption{Quantum Fisher information as a function of the asymmetry $\epsilon$. $\left(a\right)$ QFI for the attenuator channel obtained from the analytical expressions (\ref{QFIatteq}), with black and blue dots for $\varphi$ and $\bar{m}$, respectively. $\left(b\right)$ QFI for the amplification channel obtained from the analytical expressions (\ref{QFIampeq}), with black and blue dots for $r_g$ and $\bar{m}$, respectively. The straight lines represent the respective power law scaling $\mathcal{I} = \alpha + \beta\epsilon^n$, with $n = 3$ and the coefficients given by the Table \ref{table}.}
\label{QFIepsilon}
\end{figure}

To conclude the discussion about the role of coherence in the QFI, the influence of the asymmetry between the Gaussian measurements can be well captured by proposing a simple parametrization for the uncertainties, such that $\sigma_{q}=\sigma/\left(1-\epsilon\right)^{1/2}$ and $\sigma_{p}=\sigma/\left(1+\epsilon\right)^{1/2}$, where $\sigma$ controls the temperature of the probe. In contrast, the asymmetry is represented by the dimensionless factor $\epsilon$. For $\epsilon = 0$, we have basically a thermal probe. By inserting this parametrization in Eq. (\ref{aftermeasurement}), the CM immediately before any channel reads
\begin{equation}
    \Sigma^{\text{probe}} = \frac{\left(2\bar{n}+1\right)}{\sigma^{2}}\left(\mathbb{I}-\epsilon\sigma_{z}\right),
    \label{parametrization}
\end{equation}
with $\mathbb{I}$ and $\sigma_z$ the identity and Pauli matrix, respectively. It is evident from Eq. (\ref{parametrization}) that the effective contribution from the Gaussian measurements, as well as the presence of quantum coherence in the probe state, only comes in the asymmetric scenario. The effects of asymmetry in the QFI are shown in Fig.~\ref{QFIepsilon}, where we plot the QFI as a function of $\epsilon$ for different estimation parameters of the attenuator and amplification channels. Figure~\ref{QFIepsilon}-(a) shows analytical results for the beam-splitter angle $\varphi = \pi/4$ (black dots) and for the average thermal number $\bar{m} = 0.5$ (blue dots) of the attenuator channel, whereas Fig.~\ref{QFIepsilon}-(b) shows analytical results for the squeezing parameter $r_g = 1.0$ and for the average thermal number $\bar{m} = 0.5$ (blue dots) of the amplification channel. Furthermore, we also provide the power law $\mathcal{I} = \sum_n \alpha^n \epsilon^n$, with $\alpha = \alpha\left(\vec{\theta}\right)$ for each Gaussian channel. Basically, given a fixed value of the parameter to be estimated, we observe how the QFI changes with the asymmetry. Using the dots obtained from the analytical expressions (\ref{QFIatteq}) and (\ref{QFIampeq}), we find by fitting that the best power law scaling is given by the general expression $\mathcal{I} = \alpha + \beta\epsilon^n$, with $\alpha$ representing the QFI due only to thermal resources. The coefficients for each case are shown in the Table \ref{table}, and we have found the best exponent to be $n\approx 3$. 
\begin{table}[h]
    \centering
    \begin{tabular}{|c|c|}
    \hline
        \textbf{$\alpha$} & \textbf{$\beta$} \\
        \hline
        $\alpha_{\varphi=\pi/4}^{\text{att}}=0.16$ & $\beta_{\alpha=\pi/4}^{\text{att}}=1.72$ \\
        \hline
        $\alpha_{\bar{m}}^{\text{att}}=0.31$ & $\beta_{\bar{m}}^{\text{att}}=0.40$ \\
        \hline
        $\alpha_{r_{g}}^{\text{amp}}=3.72$ & $\beta^{\text{amp}}=1.30$ \\
        \hline
        $\alpha_{\bar{m}}^{\text{amp}}=0.12$ & $\beta_{\bar{m}}^{\text{amp}}=0.34$ \\
        \hline
    \end{tabular}
    \caption{Coefficients for the power law scaling for the QFI as a function of the asymmetry $\epsilon$ for the attenuator and amplification channels. For the former, we have fixed $\varphi = \pi/4$ and $\bar{m} = 0.5$, whereas for the latter we have fixed $r_g = 1.0$ and $\bar{m} = 0.5$. The probe initial parameters are the same as in Fig. \ref{QFI_att01} Fig. \ref{QFI_amp01}.}
    \label{table}
\end{table}

From an experimental point of view, power-law scaling can be verified, for instance, using quantum-optical devices, where Gaussian measurements in position and momentum can be implemented with a small asymmetry pattern. Alternatively, the asymmetry can be simulated using polarization squeezing \cite{PhysRevLett.88.093601}.

\section{Conclusion and Remarks}\label{sec03}

The precise estimation of physical parameters that represent relevant quantum processes is pivotal across different branches of quantum technologies, including quantum computation and quantum thermodynamics, as well as the development of gravimeters and magnetometers. In the present work, we focused on a particular scheme for probe-state preparation, composed of two non-commutative Gaussian measurements of the position and momentum observables. To exemplify the use of such a probe-state preparation in parameter estimation, we considered two paradigmatic, experimentally feasible single-mode Gaussian channels: the attenuator and the amplification channel.

Our results show that the use of non-commutative Gaussian measurements can improve the quantum Fisher information for parameters characterizing Gaussian channels beyond the standard limit using only a thermal probe. Our state preparation protocol corresponds to two classes: the symmetric and the asymmetric scenarios, in which only the asymmetric scenario introduces quantum coherence in the energy basis of the probe state. We also quantified the derivative of the coherence with respect to the parameters to be estimated, showing that its rate of change is directly associated with the improvement or decrease of the QFI, evidencing that when the derivative is zero, the QFI assumes a constant value.

To conclude, our results are experimentally feasible, with the Gaussian channels employed here simulating relevant physical processes, such as beam-splitter and squeezing. A possible implementation is based on optical devices that emulate canonical variables in light's polarization modes. We hope that our present work helps to elucidate the role of coherence and its origins in quantum metrology protocols. Possible future directions include, for instance, the use of non-commutative Gaussian measurements in the preparation of probes for gravimetry and magnometry, as well as the construction of non-commutative measurements with observables other than the canonical position and momentum ones.

\begin{acknowledgments}
Alice P. G. Hall acknowledges Universidade Federal da Grande Dourados
for support. Carlos H. S. Vieira acknowledges the São Paulo Research Foundation (FAPESP) Grant No. 2023/13362-0 and Grant No. 2025/14546-2 for financial support and the Southern University of Science and Technology (SUSTech) for providing the workspace. Jonas F. G. Santos acknowledges CNPq Grant No. 420549/2023-4, Fundect
Grant No. 83/026.973/2024, and Universidade Federal da Grande Dourados
for support.
\end{acknowledgments}

\section*{Appendix - Action of the Gaussian measurements}\label{Appendix}

We here show in detail the derivation of the covariance matrix of the probe state after the two non-commuting Gaussian measurements, Eqs. (\ref{aftermeasurement}).  Before the measurements, the probe is initially prepared in a thermal state, such that in the Fock basis it is
\begin{equation}
    \rho_{\text{th}}  =\sum_{n=0}^{\infty}\frac{\bar{n}^{n}}{\left(\bar{n}+1\right)^{n+1}}|n\rangle\langle n|,
\end{equation}
with $\bar{n} = \text{Tr}\left[\rho_{\text{th}} a^\dagger a\right]$.  The first moments are zero for a thermal state, and they are not affected by the measurements.

Let us compute the elements of the covariance matrix. The first measurement is in the position basis, with $r_1 = q$ in Eq. (\ref{Gaussianmea}), such that the post-first measurement state is formally written as $\rho^{M_q}=\int d\alpha_q M_{\alpha_q}\rho^{th}M_{\alpha_q}^{\dagger}$. The measurement operator commutes with the operator $q$, such that $\Sigma^{M_q}_{11}= 2\langle q^{2}\rangle_{\rho^{M}}-2\langle q\rangle_{\rho^{M}}^{2}$, and $\Sigma^{M_q}_{11} = 2\langle q^{2}\rangle_{\rho_{\text{th}}}  = \left(\hbar/m\omega\right)\left(2\bar{n} + 1\right)$. On the other hand, for the momentum operator, $\Sigma^{M_q}_{22}= 2\langle p^{2}\rangle_{\rho^{M}}-2\langle p\rangle_{\rho^{M}}^{2}$, with the second term identically zero. Then,
\begin{equation}
    \Sigma^{M_q}_{22}=2\int d\alpha Tr\left[M_{\alpha}^{\dagger}p^{2}M_{\alpha}\rho^{th}\right],
\end{equation}
with $p^2 = -\hbar^2\partial^2/\partial q^2$ in the position basis. By expanding the derivative and using the results
\begin{eqnarray}
    \int d\alpha\exp\left[-\frac{\left(q-\alpha\right)^{2}}{2\sigma_{q}^{2}}\right]&=\sqrt{2\pi}\sigma_{q},\nonumber\\
    \int d\alpha\left(q-\alpha\right)^{2}\exp\left[-\frac{\left(q-\alpha\right)^{2}}{2\sigma_{q}^{2}}\right]&=\sqrt{2\pi}\sigma_{q}^{3},\nonumber
\end{eqnarray}
we obtain $\Sigma_{22}^{M_q} = \hbar^2/\left(2\sigma_q^2\right)$. By analogy, the elements $\sigma_{12}^{M_q}  = \sigma_{21}^{M_q}  = 0$.

In the sequence, we measure in the momentum basis. The calculation is identical, except that now the measurement operator commutes with the element $\sigma_{22}^{M_q}$. The elements give the covariance matrix after the two measurements
\begin{eqnarray}
    \sigma_{11}^{M_qM_p} &=&   \frac{\hbar\left(2\bar{n}+1\right)}{2m\omega\sigma_{p}^{2}}  ,\nonumber \\
    \sigma_{12}^{M_qM_p} &=& 0    ,\nonumber\\
    \sigma_{21}^{M_qM_p} &=& 0  ,\nonumber\\
    \sigma_{22}^{M_qM_p} &=& \frac{\hbar m\omega}{2}\frac{\left(2\bar{n}+1\right)}{\sigma_{q}^{2}} ,\nonumber
\end{eqnarray}
where we did the parametrization $\sigma_{q}^{2}\rightarrow\frac{\hbar}{m\omega}\frac{\sigma_{q}^{2}}{2\bar{n}+1}$  and $\sigma_{p}^{2} \rightarrow \frac{\hbar m\omega}{\left(2\bar{n}+1\right)}\sigma_{p}^{2}$, such that $\sigma_q = \sigma_p = 1$ represents the scenario in which the measurements do nothing on the initial probe state.

\bibliography{Refs}

@article{PhysRevE.96.022108,
  title = {Single-temperature quantum engine without feedback control},
  author = {Yi, Juyeon and Talkner, Peter and Kim, Yong Woon},
  journal = {Phys. Rev. E},
  volume = {96},
  issue = {2},
  pages = {022108},
  numpages = {5},
  year = {2017},
  month = {Aug},
  publisher = {American Physical Society},
  doi = {10.1103/PhysRevE.96.022108},
  url = {https://link.aps.org/doi/10.1103/PhysRevE.96.022108}
}

@article{Huang_2024,
author = {Huang, Jiahao and Zhuang, Min and Lee, Chaohong},
title = "{Entanglement-enhanced quantum metrology: From standard quantum limit to Heisenberg limit}",
journal = {Applied Physics Reviews},
volume = {11},
number = {3},
pages = {031302},
year = {2024},
month = {07},
issn = {1931-9401},
doi = {10.1063/5.0204102},
url = {https://doi.org/10.1063/5.0204102},
}

@article{PARIS_09,
author = {Paris, Matteo G. A.},
title = {Quantum Estimation For Quantum Technology},
journal = {International Journal of Quantum Information},
volume = {07},
number = {supp01},
pages = {125-137},
year = {2009},
doi = {10.1142/S0219749909004839},
URL = {https://doi.org/10.1142/S0219749909004839}
}

@article{Liu_2020,
doi = {10.1088/1751-8121/ab5d4d},
url = {https://dx.doi.org/10.1088/1751-8121/ab5d4d},
year = {2019},
month = {dec},
publisher = {IOP Publishing},
volume = {53},
number = {2},
pages = {023001},
author = {Jing Liu and Haidong Yuan and Xiao-Ming Lu and Xiaoguang Wang},
title = {Quantum Fisher information matrix and multiparameter estimation},
journal = {Journal of Physics A: Mathematical and Theoretical}
}

@article{Giovannetti_nat18,
author={Rosati, Matteo
and Mari, Andrea
and Giovannetti, Vittorio},
title={Narrow bounds for the quantum capacity of thermal attenuators},
journal={Nature Communications},
year={2018},
month={Oct},
day={18},
volume={9},
number={1},
pages={4339},
issn={2041-1723},
doi={10.1038/s41467-018-06848-0},
url={https://doi.org/10.1038/s41467-018-06848-0}
}

@article{Giovannetti_prl21,
  title = {{Estimating Quantum and Private Capacities of Gaussian Channels via Degradable Extensions}},
  author = {Fanizza, Marco and Kianvash, Farzad and Giovannetti, Vittorio},
  journal = {Phys. Rev. Lett.},
  volume = {127},
  issue = {21},
  pages = {210501},
  numpages = {6},
  year = {2021},
  month = {Nov},
  publisher = {American Physical Society},
  doi = {10.1103/PhysRevLett.127.210501},
  url = {https://link.aps.org/doi/10.1103/PhysRevLett.127.210501}
}

@article{PhysRevA.88.040102,
  title = {{Quantum parameter estimation using general single-mode Gaussian states}},
  author = {Pinel, O. and Jian, P. and Treps, N. and Fabre, C. and Braun, D.},
  journal = {Phys. Rev. A},
  volume = {88},
  issue = {4},
  pages = {040102},
  numpages = {5},
  year = {2013},
  month = {Oct},
  publisher = {American Physical Society},
  doi = {10.1103/PhysRevA.88.040102},
  url = {https://link.aps.org/doi/10.1103/PhysRevA.88.040102}
}

@article{Fuentes2015,
  title = {{Quantum parameter estimation using multi-mode Gaussian states}},
  author = {D. Safranek and A. R. Lee and I. Fuentes},
  journal = {New J. Phys.},
  volume = {17},
  pages = {073016},
  year = {2015},
  doi = {10.1088/1367-2630/17/7/073016},
  url = {https://iopscience.iop.org/article/10.1088/1367-2630/17/7/073016}
}

@article{RevModPhys.84.621,
  title = {{Gaussian quantum information}},
  author = {Weedbrook, Christian and Pirandola, Stefano and Garc\'{\i}a-Patr\'on, Ra\'ul and Cerf, Nicolas J. and Ralph, Timothy C. and Shapiro, Jeffrey H. and Lloyd, Seth},
  journal = {Rev. Mod. Phys.},
  volume = {84},
  issue = {2},
  pages = {621--669},
  numpages = {0},
  year = {2012},
  month = {May},
  publisher = {American Physical Society},
  doi = {10.1103/RevModPhys.84.621},
  url = {https://link.aps.org/doi/10.1103/RevModPhys.84.621}
}

@article{Baumgratz2014,
  title = {Quantifying Coherence},
  author = {Baumgratz, T. and Cramer, M. and Plenio, M. B.},
  journal = {Phys. Rev. Lett.},
  volume = {113},
  issue = {14},
  pages = {140401},
  numpages = {5},
  year = {2014},
  month = {Sep},
  publisher = {American Physical Society},
  doi = {10.1103/PhysRevLett.113.140401},
  url = {https://link.aps.org/doi/10.1103/PhysRevLett.113.140401}
}

@article{Streltsov2017,
  title = {Colloquium: Quantum coherence as a resource},
  author = {Streltsov, Alexander and Adesso, Gerardo and Plenio, Martin B.},
  journal = {Rev. Mod. Phys.},
  volume = {89},
  issue = {4},
  pages = {041003},
  numpages = {34},
  year = {2017},
  month = {Oct},
  publisher = {American Physical Society},
  doi = {10.1103/RevModPhys.89.041003},
  url = {https://link.aps.org/doi/10.1103/RevModPhys.89.041003}
}

@article{Xu2016,
  title = {{Quantifying coherence of Gaussian states}},
  author = {Xu, Jianwei},
  journal = {Phys. Rev. A},
  volume = {93},
  issue = {3},
  pages = {032111},
  numpages = {7},
  year = {2016},
  month = {Mar},
  publisher = {American Physical Society},
  doi = {10.1103/PhysRevA.93.032111},
  url = {https://link.aps.org/doi/10.1103/PhysRevA.93.032111}
}

@article{eisert2005gaussian,
  title={Gaussian quantum channels},
  author={Eisert, Jens and Wolf, Michael M},
  journal={arXiv preprint quant-ph/0505151},
  year={2005},
  url={
https://doi.org/10.48550/arXiv.quant-ph/0505151
}
}

@book{serafini2017quantum,
  title={Quantum continuous variables: a primer of theoretical methods},
  author={Serafini, Alessio},
  year={2017},
  publisher={CRC press},
  url={https://doi.org/10.1201/9781315118727}
}

@article{Cenni2022thermometryof,
  doi = {10.22331/q-2022-06-23-743},
  url = {https://doi.org/10.22331/q-2022-06-23-743},
  title = {Thermometry of {G}aussian quantum systems using {G}aussian measurements},
  author = {Cenni, Marina F.B. and Lami, Ludovico and Ac{\'{i}}n, Antonio and Mehboudi, Mohammad},
  journal = {{Quantum}},
  issn = {2521-327X},
  publisher = {{Verein zur F{\"{o}}rderung des Open Access Publizierens in den Quantenwissenschaften}},
  volume = {6},
  pages = {743},
  month = jun,
  year = {2022}
}

@article{f79z-vjsb,
  title = {Achieving the Heisenberg limit of metrology via measurement on an ancillary qubit},
  author = {Chen, Peng and Jing, Jun},
  journal = {Phys. Rev. A},
  volume = {112},
  issue = {3},
  pages = {032416},
  numpages = {12},
  year = {2025},
  month = {Sep},
  publisher = {American Physical Society},
  doi = {10.1103/f79z-vjsb},
  url = {https://link.aps.org/doi/10.1103/f79z-vjsb}
}

@article{PhysRevA.104.042609,
  title = {Gaussian quantum metrology in a dissipative environment},
  author = {Wu, Wei and An, Jun-Hong},
  journal = {Phys. Rev. A},
  volume = {104},
  issue = {4},
  pages = {042609},
  numpages = {7},
  year = {2021},
  month = {Oct},
  publisher = {American Physical Society},
  doi = {10.1103/PhysRevA.104.042609},
  url = {https://link.aps.org/doi/10.1103/PhysRevA.104.042609}
}

@article{PhysRevLett.131.150803,
  title = {Generalized Quantum Measurements on a Higher-Dimensional System via Quantum Walks},
  author = {Wang, Xiaowei and Zhan, Xiang and Li, Yulin and Xiao, Lei and Zhu, Gaoyan and Qu, Dengke and Lin, Quan and Yu, Yue and Xue, Peng},
  journal = {Phys. Rev. Lett.},
  volume = {131},
  issue = {15},
  pages = {150803},
  numpages = {7},
  year = {2023},
  month = {Oct},
  publisher = {American Physical Society},
  doi = {10.1103/PhysRevLett.131.150803},
  url = {https://link.aps.org/doi/10.1103/PhysRevLett.131.150803}
}

@article{Santos2025,
  title = {Improving parameter estimation in Gaussian channels using quantum coherence},
  author = {Santos, Jonas F. G. and Vieira, Carlos H. S. and Cardoso, Wilder R.},
  journal = {Phys. Rev. A},
  volume = {111},
  issue = {5},
  pages = {052404},
  numpages = {14},
  year = {2025},
  month = {May},
  publisher = {American Physical Society},
  doi = {10.1103/PhysRevA.111.052404},
  url = {https://link.aps.org/doi/10.1103/PhysRevA.111.052404}
}

@article{PhysRevA.98.012114,
  title = {Multiparameter Gaussian quantum metrology},
  author = {Nichols, Rosanna and Liuzzo-Scorpo, Pietro and Knott, Paul A. and Adesso, Gerardo},
  journal = {Phys. Rev. A},
  volume = {98},
  issue = {1},
  pages = {012114},
  numpages = {13},
  year = {2018},
  month = {Jul},
  publisher = {American Physical Society},
  doi = {10.1103/PhysRevA.98.012114},
  url = {https://link.aps.org/doi/10.1103/PhysRevA.98.012114}
}

@article{giovannetti2011advances,
  title={Advances in quantum metrology},
  author={Giovannetti, Vittorio and Lloyd, Seth and Maccone, Lorenzo},
  journal={Nature photonics},
  volume={5},
  number={4},
  pages={222--229},
  year={2011},
  publisher={Nature Publishing Group UK London},
doi={https://doi.org/10.1038/nphoton.2011.35}
}

@article{RevModPhys.90.035005,
  title = {Quantum metrology with nonclassical states of atomic ensembles},
  author = {Pezz\`e, Luca and Smerzi, Augusto and Oberthaler, Markus K. and Schmied, Roman and Treutlein, Philipp},
  journal = {Rev. Mod. Phys.},
  volume = {90},
  issue = {3},
  pages = {035005},
  numpages = {70},
  year = {2018},
  month = {Sep},
  publisher = {American Physical Society},
  doi = {10.1103/RevModPhys.90.035005},
  url = {https://link.aps.org/doi/10.1103/RevModPhys.90.035005}
}

@article{schnabel2010quantum,
  title={Quantum metrology for gravitational wave astronomy},
  author={Schnabel, Roman and Mavalvala, Nergis and McClelland, David E and Lam, Ping K},
  journal={Nature communications},
  volume={1},
  number={1},
  pages={121},
  year={2010},
  publisher={Nature Publishing Group UK London}
}

@article{toth2014quantum,
  title={Quantum metrology from a quantum information science perspective},
  author={T{\'o}th, G{\'e}za and Apellaniz, Iagoba},
  journal={Journal of Physics A: Mathematical and Theoretical},
  volume={47},
  number={42},
  pages={424006},
  year={2014},
  publisher={IOP Publishing}
}

@article{haase2016precision,
  title={Precision limits in quantum metrology with open quantum systems},
  author={Haase, Jan F and Smirne, Andrea and Huelga, SF and Ko{\l}odynski, J and Demkowicz-Dobrzanski, R},
  journal={Quantum Measurements and Quantum Metrology},
  volume={5},
  number={1},
  pages={13--39},
  year={2016},
  publisher={De Gruyter Open}
}

@article{zwierz2012ultimate,
  title={Ultimate limits to quantum metrology and the meaning of the Heisenberg limit},
  author={Zwierz, Marcin and P{\'e}rez-Delgado, Carlos A and Kok, Pieter},
  journal={Physical Review A—Atomic, Molecular, and Optical Physics},
  volume={85},
  number={4},
  pages={042112},
  year={2012},
  publisher={APS}
}

@article{seveso2017quantum,
  title={Quantum metrology beyond the quantum Cram{\'e}r-Rao theorem},
  author={Seveso, Luigi and Rossi, Matteo AC and Paris, Matteo GA},
  journal={Physical Review A},
  volume={95},
  number={1},
  pages={012111},
  year={2017},
  publisher={APS}
}

@article{PhysRevLett.133.120601,
  title = {Modular Many-Body Quantum Sensors},
  author = {Mukhopadhyay, Chiranjib and Bayat, Abolfazl},
  journal = {Phys. Rev. Lett.},
  volume = {133},
  issue = {12},
  pages = {120601},
  numpages = {7},
  year = {2024},
  month = {Sep},
  publisher = {American Physical Society},
  doi = {10.1103/PhysRevLett.133.120601},
  url = {https://link.aps.org/doi/10.1103/PhysRevLett.133.120601}
}

@article{mukhopadhyay2024current,
  title={Current trends in global quantum metrology},
  author={Mukhopadhyay, Chiranjib and Montenegro, Victor and Bayat, Abolfazl},
  journal={Journal of Physics A: Mathematical and Theoretical},
  year={2024}
}

@article{montenegro2024review,
  title={Review: Quantum metrology and sensing with many-body systems},
  author={Montenegro, Victor and Mukhopadhyay, Chiranjib and Yousefjani, Rozhin and Sarkar, Saubhik and Mishra, Utkarsh and Paris, Matteo GA and Bayat, Abolfazl},
  journal={arXiv preprint arXiv:2408.15323},
  year={2024}
}

@article{morelli2021bayesian,
  title={Bayesian parameter estimation using Gaussian states and measurements},
  author={Morelli, Simon and Usui, Ayaka and Agudelo, Elizabeth and Friis, Nicolai},
  journal={Quantum Science and Technology},
  volume={6},
  number={2},
  pages={025018},
  year={2021},
  publisher={IOP Publishing}
}

@article{Pritam2024QST,
doi = {10.1088/2058-9565/adcae3},
url = {https://dx.doi.org/10.1088/2058-9565/adcae3},
year = {2025},
month = {apr},
publisher = {IOP Publishing},
volume = {10},
number = {3},
pages = {035006},
author = {Chattopadhyay, Pritam and Misra, Avijit and Sur, Saikat and Petrosyan, David and Kurizki, Gershon},
title = {Sensing multiatom networks in cavities via photon-induced excitation resonance},
journal = {Quantum Science and Technology}
}

@article{di2023critical,
  title={Critical parametric quantum sensing},
  author={Di Candia, Roberto and Minganti, Fabrizio and Petrovnin, KV and Paraoanu, Gheorghe S and Felicetti, Simone},
  journal={npj Quantum Information},
  volume={9},
  number={1},
  pages={23},
  year={2023},
  publisher={Nature Publishing Group UK London}
}

@article{petrovnin2024microwave,
  title={Microwave photon detection at parametric criticality},
  author={Petrovnin, Kirill and Wang, Jiaming and Perelshtein, Michael and Hakonen, Pertti and Paraoanu, Gheorghe Sorin},
  journal={PRX Quantum},
  volume={5},
  number={2},
  pages={020342},
  year={2024},
  publisher={APS}
}

@article{degen2017quantum,
  title={Quantum sensing},
  author={Degen, Christian L and Reinhard, Friedemann and Cappellaro, Paola},
  journal={Reviews of modern physics},
  volume={89},
  number={3},
  pages={035002},
  year={2017},
  publisher={APS}
}

@ARTICLE{Chattopadhyay2025-jk,
  title     = "Generic two-mode Gaussian states as quantum sensors",
  author    = "Chattopadhyay, Pritam and Sur, Saikat and Santos, Jonas F G",
  journal   = "Quantum Sci. Technol.",
  publisher = "IOP Publishing",
  volume    =  10,
  number    =  4,
  pages     = "045044",
  month     =  dec,
  year      =  2025,
  copyright = "https://publishingsupport.iopscience.iop.org/iop-standard/v1"
}

@article{Giovannetti2004-el,
  title     = "Quantum-enhanced measurements: beating the standard quantum
               limit",
  author    = "Giovannetti, Vittorio and Lloyd, Seth and Maccone, Lorenzo",
  journal   = "Science",
  publisher = "American Association for the Advancement of Science (AAAS)",
  volume    =  306,
  number    =  5700,
  pages     = "1330--1336",
  month     =  nov,
  year      =  2004,
}

@article{PhysRevLett.96.010401,
  title = {Quantum Metrology},
  author = {Giovannetti, Vittorio and Lloyd, Seth and Maccone, Lorenzo},
  journal = {Phys. Rev. Lett.},
  volume = {96},
  issue = {1},
  pages = {010401},
  numpages = {4},
  year = {2006},
  month = {Jan},
  publisher = {American Physical Society},
  doi = {10.1103/PhysRevLett.96.010401},
  url = {https://link.aps.org/doi/10.1103/PhysRevLett.96.010401}
}

@ARTICLE{Daryanoosh2018-nn,
  title     = "Experimental optical phase measurement approaching the exact
               Heisenberg limit",
  author    = "Daryanoosh, Shakib and Slussarenko, Sergei and Berry, Dominic W
               and Wiseman, Howard M and Pryde, Geoff J",
  journal   = "Nat. Commun.",
  publisher = "Springer Science and Business Media LLC",
  volume    =  9,
  number    =  1,
  pages     = "4606",
  month     =  nov,
  year      =  2018,
  copyright = "https://creativecommons.org/licenses/by/4.0",
}

@article{PhysRevA.110.012614,
  title = {Experimental investigation of a multiphoton Heisenberg-limited interferometric scheme: The effect of imperfections},
  author = {Daryanoosh, Shakib and Pryde, Geoff J. and Wiseman, Howard M. and Slussarenko, Sergei},
  journal = {Phys. Rev. A},
  volume = {110},
  issue = {1},
  pages = {012614},
  numpages = {16},
  year = {2024},
  month = {Jul},
  publisher = {American Physical Society},
  doi = {10.1103/PhysRevA.110.012614},
  url = {https://link.aps.org/doi/10.1103/PhysRevA.110.012614}
}

@ARTICLE{DeAraujo2025ws,
  title         = "Towards gravimetry enhancement with squeezed states",
  author        = "de Araujo, Oziel R and Marinho, Lucas S and Santos, Jonas F
                   G and Vieira, Carlos H S",
  journal   = "Arxiv",
  month         =  oct,
  year          =  2025,
  copyright     = "http://arxiv.org/licenses/nonexclusive-distrib/1.0/",
  archivePrefix = "arXiv",
  primaryClass  = "quant-ph",
  eprint        = "2510.13973"
}

@ARTICLE{Qvarfort2018-dr,
  title     = "Gravimetry through non-linear optomechanics",
  author    = "Qvarfort, Sofia and Serafini, Alessio and Barker, P F and Bose,
               Sougato",
  journal   = "Nat. Commun.",
  publisher = "Springer Science and Business Media LLC",
  volume    =  9,
  number    =  1,
  pages     = "3690",
  month     =  sep,
  year      =  2018,
  copyright = "https://creativecommons.org/licenses/by/4.0"
}

@article{PhysRevLett.117.138501,
  title = {Simultaneous Precision Gravimetry and Magnetic Gradiometry with a Bose-Einstein Condensate: A High Precision, Quantum Sensor},
  author = {Hardman, K. S. and Everitt, P. J. and McDonald, G. D. and Manju, P. and Wigley, P. B. and Sooriyabandara, M. A. and Kuhn, C. C. N. and Debs, J. E. and Close, J. D. and Robins, N. P.},
  journal = {Phys. Rev. Lett.},
  volume = {117},
  issue = {13},
  pages = {138501},
  numpages = {5},
  year = {2016},
  month = {Sep},
  publisher = {American Physical Society},
  doi = {10.1103/PhysRevLett.117.138501},
  url = {https://link.aps.org/doi/10.1103/PhysRevLett.117.138501}
}

@article{PhysRevResearch.7.013016,
  title = {Heisenberg-limited spin-mechanical gravimetry},
  author = {Montenegro, Victor},
  journal = {Phys. Rev. Res.},
  volume = {7},
  issue = {1},
  pages = {013016},
  numpages = {15},
  year = {2025},
  month = {Jan},
  publisher = {American Physical Society},
  doi = {10.1103/PhysRevResearch.7.013016},
  url = {https://link.aps.org/doi/10.1103/PhysRevResearch.7.013016}
}

@ARTICLE{Boixo2009-yw,
  title     = "Quantum-limited metrology and {Bose-Einstein} condensates",
  author    = "Boixo, Sergio and Datta, Animesh and Davis, Matthew J and Shaji,
               Anil and Tacla, Alexandre B and Caves, Carlton M",
  journal   = "Phys. Rev. A",
  publisher = "American Physical Society (APS)",
  volume    =  80,
  number    =  3,
  month     =  sep,
  year      =  2009,
  copyright = "http://link.aps.org/licenses/aps-default-license"
}

@ARTICLE{Ngo2021-iy,
  title     = "{Bose-Einstein} condensate soliton qubit states for metrological
               applications",
  author    = "Ngo, The Vinh and Tsarev, Dmitriy V and Lee, Ray-Kuang and
               Alodjants, Alexander P",
  journal   = "Sci. Rep.",
  publisher = "Springer Science and Business Media LLC",
  volume    =  11,
  number    =  1,
  pages     = "19363",
  month     =  sep,
  year      =  2021,
  copyright = "https://creativecommons.org/licenses/by/4.0"
}

@ARTICLE{Dalvit2006-cq,
  title     = "Quantum metrology at the Heisenberg limit with ion trap motional
               compass states",
  author    = "Dalvit, D A R and Filho, R L de Matos and Toscano, F",
  journal   = "New J. Phys.",
  publisher = "IOP Publishing",
  volume    =  8,
  number    =  11,
  pages     = "276--276",
  month     =  nov,
  year      =  2006
}

@ARTICLE{Delakouras2024-lr,
  title     = "Production of Fock mixtures in trapped ions for motional
               metrology",
  author    = "Delakouras, Antonis and Rodr{\'\i}guez, Daniel and Cerrillo,
               Javier",
  journal   = "Quantum Sci. Technol.",
  publisher = "IOP Publishing",
  volume    =  9,
  number    =  1,
  pages     = "015006",
  month     =  jan,
  year      =  2024,
  copyright = "https://iopscience.iop.org/page/copyright"
}

@ARTICLE{Frerot2024-pi,
  title     = "Symmetry: A fundamental resource for quantum coherence and
               metrology",
  author    = "Fr{\'e}rot, Ir{\'e}n{\'e}e and Roscilde, Tommaso",
  journal   = "Phys. Rev. Lett.",
  publisher = "American Physical Society (APS)",
  volume    =  133,
  number    =  26,
  pages     = "260402",
  month     =  dec,
  year      =  2024,
  copyright = "https://link.aps.org/licenses/aps-default-license",
}

@ARTICLE{Lecamwasam2024-tn,
  title     = "Relative entropy of coherence quantifies performance in Bayesian
               metrology",
  author    = "Lecamwasam, Ruvi and Assad, Syed and Hope, Joseph J and Lam,
               Ping Koy and Thompson, Jayne and Gu, Mile",
  journal   = "PRX quantum",
  publisher = "American Physical Society (APS)",
  volume    =  5,
  number    =  3,
  month     =  jul,
  year      =  2024,
  copyright = "https://creativecommons.org/licenses/by/4.0/",
}

@ARTICLE{Deng2024-uj,
  title     = "Quantum-enhanced metrology with large Fock states",
  author    = "Deng, Xiaowei and Li, Sai and Chen, Zi-Jie and Ni, Zhongchu and
               Cai, Yanyan and Mai, Jiasheng and Zhang, Libo and Zheng, Pan and
               Yu, Haifeng and Zou, Chang-Ling and Liu, Song and Yan, Fei and
               Xu, Yuan and Yu, Dapeng",
  journal   = "Nat. Phys.",
  publisher = "Springer Science and Business Media LLC",
  volume    =  20,
  number    =  12,
  pages     = "1874--1880",
  month     =  dec,
  year      =  2024,
  copyright = "https://creativecommons.org/licenses/by/4.0",
}

@article{PhysRevLett.134.180801,
  title = {Genuine Quantum Non-Gaussianity and Metrological Sensitivity of Fock States Prepared in a Mechanical Resonator},
  author = {Rahman, Q. Rumman and Kladari\ifmmode \acute{c}\else \'{c}\fi{}, Igor and Kern, Max-Emanuel and Lachman, Luk\'a\ifmmode \check{s}\else \v{s}\fi{} and Chu, Yiwen and Filip, Radim and Fadel, Matteo},
  journal = {Phys. Rev. Lett.},
  volume = {134},
  issue = {18},
  pages = {180801},
  numpages = {6},
  year = {2025},
  month = {May},
  publisher = {American Physical Society},
  doi = {10.1103/PhysRevLett.134.180801},
  url = {https://link.aps.org/doi/10.1103/PhysRevLett.134.180801}
}

@article{PhysRevA.104.062210,
  title = {Suppressing coherence effects in quantum-measurement-based engines},
  author = {Lin, Zhiyuan and Su, Shanhe and Chen, Jingyi and Chen, Jincan and Santos, Jonas F. G.},
  journal = {Phys. Rev. A},
  volume = {104},
  issue = {6},
  pages = {062210},
  numpages = {8},
  year = {2021},
  month = {Dec},
  publisher = {American Physical Society},
  doi = {10.1103/PhysRevA.104.062210},
  url = {https://link.aps.org/doi/10.1103/PhysRevA.104.062210}
}

@article{PhysRevE.110.L052102,
  title = {Finite-time measurement-driven Otto cycle},
  author = {Wang, Youlin and Xia, Shihao and Lin, Xinqiao and Pan, Ousi and Chen, Jincan and Su, Shanhe},
  journal = {Phys. Rev. E},
  volume = {110},
  issue = {5},
  pages = {L052102},
  numpages = {7},
  year = {2024},
  month = {Nov},
  publisher = {American Physical Society},
  doi = {10.1103/PhysRevE.110.L052102},
  url = {https://link.aps.org/doi/10.1103/PhysRevE.110.L052102}
}

@article{PhysRevLett.126.120605,
  title = {Two-Qubit Engine Fueled by Entanglement and Local Measurements},
  author = {Bresque, L\'ea and Camati, Patrice A. and Rogers, Spencer and Murch, Kater and Jordan, Andrew N. and Auff\`eves, Alexia},
  journal = {Phys. Rev. Lett.},
  volume = {126},
  issue = {12},
  pages = {120605},
  numpages = {6},
  year = {2021},
  month = {Mar},
  publisher = {American Physical Society},
  doi = {10.1103/PhysRevLett.126.120605},
  url = {https://link.aps.org/doi/10.1103/PhysRevLett.126.120605}
}

@article{PhysRevLett.120.260601,
  title = {Efficient Quantum Measurement Engines},
  author = {Elouard, Cyril and Jordan, Andrew N.},
  journal = {Phys. Rev. Lett.},
  volume = {120},
  issue = {26},
  pages = {260601},
  numpages = {5},
  year = {2018},
  month = {Jun},
  publisher = {American Physical Society},
  doi = {10.1103/PhysRevLett.120.260601},
  url = {https://link.aps.org/doi/10.1103/PhysRevLett.120.260601}
}

@article{PhysRevResearch.7.013151,
  title = {Nonequilibrium quantum battery based on quantum measurements},
  author = {Du, Jianying and Guo, Yanjiang and Li, Baowen},
  journal = {Phys. Rev. Res.},
  volume = {7},
  issue = {1},
  pages = {013151},
  numpages = {15},
  year = {2025},
  month = {Feb},
  publisher = {American Physical Society},
  doi = {10.1103/PhysRevResearch.7.013151},
  url = {https://link.aps.org/doi/10.1103/PhysRevResearch.7.013151}
}

@article{PhysRevA.109.042424,
  title = {Local-projective-measurement-enhanced quantum battery capacity},
  author = {Zhang, Tinggui and Yang, Hong and Fei, Shao-Ming},
  journal = {Phys. Rev. A},
  volume = {109},
  issue = {4},
  pages = {042424},
  numpages = {6},
  year = {2024},
  month = {Apr},
  publisher = {American Physical Society},
  doi = {10.1103/PhysRevA.109.042424},
  url = {https://link.aps.org/doi/10.1103/PhysRevA.109.042424}
}

@ARTICLE{Lustosa2025-mr,
  title     = "Emergence of realism through quantum discord suppression in
               photonic weak measurements",
  author    = "Lustosa, Fabr{\'\i}cio and Barreto, Diego G and Lima, Eduardo C
               and Cruz, Luciano S and Dieguez, Pedro R and Marques, Breno",
  journal   = "New J. Phys.",
  publisher = "IOP Publishing",
  volume    =  27,
  number    =  5,
  pages     = "054503",
  month     =  may,
  year      =  2025,
  copyright = "https://creativecommons.org/licenses/by/4.0/"
}

@article{PhysRevLett.116.080403,
  title = {Thermodynamics of Weakly Measured Quantum Systems},
  author = {Alonso, Jose Joaquin and Lutz, Eric and Romito, Alessandro},
  journal = {Phys. Rev. Lett.},
  volume = {116},
  issue = {8},
  pages = {080403},
  numpages = {6},
  year = {2016},
  month = {Feb},
  publisher = {American Physical Society},
  doi = {10.1103/PhysRevLett.116.080403},
  url = {https://link.aps.org/doi/10.1103/PhysRevLett.116.080403}
}

@ARTICLE{Ballesteros_Ferraz2024-mk,
  title     = "On the relevance of weak measurements in dissipative quantum
               systems",
  author    = "Ballesteros Ferraz, Lorena and Martin, John and Caudano, Yves",
  journal   = "Quantum Sci. Technol.",
  publisher = "IOP Publishing",
  volume    =  9,
  number    =  3,
  pages     = "035029",
  month     =  jul,
  year      =  2024,
  copyright = "https://iopscience.iop.org/page/copyright"
}

@article{PhysRevA.106.022436,
  title = {Experimental investigation of a quantum heat engine powered by generalized measurements},
  author = {Lisboa, V. F. and Dieguez, P. R. and Guimar\~aes, J. R. and Santos, J. F. G. and Serra, R. M.},
  journal = {Phys. Rev. A},
  volume = {106},
  issue = {2},
  pages = {022436},
  numpages = {8},
  year = {2022},
  month = {Aug},
  publisher = {American Physical Society},
  doi = {10.1103/PhysRevA.106.022436},
  url = {https://link.aps.org/doi/10.1103/PhysRevA.106.022436}
}

@article{PhysRevA.107.012423,
  title = {Thermal devices powered by generalized measurements with indefinite causal order},
  author = {Dieguez, Pedro R. and Lisboa, Vinicius F. and Serra, Roberto M.},
  journal = {Phys. Rev. A},
  volume = {107},
  issue = {1},
  pages = {012423},
  numpages = {13},
  year = {2023},
  month = {Jan},
  publisher = {American Physical Society},
  doi = {10.1103/PhysRevA.107.012423},
  url = {https://link.aps.org/doi/10.1103/PhysRevA.107.012423}
}

@article{PhysRevLett.88.093601,
  title = {Polarization Squeezing of Continuous Variable Stokes Parameters},
  author = {Bowen, Warwick P. and Schnabel, Roman and Bachor, Hans-A. and Lam, Ping Koy},
  journal = {Phys. Rev. Lett.},
  volume = {88},
  issue = {9},
  pages = {093601},
  numpages = {4},
  year = {2002},
  month = {Feb},
  publisher = {American Physical Society},
  doi = {10.1103/PhysRevLett.88.093601},
  url = {https://link.aps.org/doi/10.1103/PhysRevLett.88.093601}
}

@article{PhysRevLett.108.253601,
  title = {Ancilla-Assisted Calibration of a Measuring Apparatus},
  author = {Brida, G. and Ciavarella, L. and Degiovanni, I. P. and Genovese, M. and Migdall, A. and Mingolla, M. G. and Paris, M. G. A. and Piacentini, F. and Polyakov, S. V.},
  journal = {Phys. Rev. Lett.},
  volume = {108},
  issue = {25},
  pages = {253601},
  numpages = {5},
  year = {2012},
  month = {Jun},
  publisher = {American Physical Society},
  doi = {10.1103/PhysRevLett.108.253601},
  url = {https://link.aps.org/doi/10.1103/PhysRevLett.108.253601}
}

\end{document}